\documentclass[a4paper,11pt]{article}
\usepackage{graphicx} 
\usepackage[
backend=biber,
style=numeric-comp,
sorting=none,
maxbibnames=99]{biblatex}
\usepackage{epsfig,latexsym,amsfonts,amsmath,amsthm,amssymb,amsbsy,multirow,slashed,wasysym,textcomp,dsfont, comment,mathtools,cancel,diagbox,datetime,appendix,BOONDOX-calo}

\usepackage{hyperref}
\usepackage{subcaption}

\usepackage{mathrsfs} %
\usepackage{calligra} %
\usepackage{braket} %
\usepackage{slashed} %
\usepackage{cancel}

\newcommand{\badat}{\begin{alignedat}}
\newcommand{\eadat}{\end{alignedat}}

\usepackage{color}

\newcommand{\Acal}{\mathcal{A}}
\newcommand{\Bcal}{\mathcal{B}}
\newcommand{\Ccal}{\mathcal{C}}
\newcommand{\Dcal}{\mathcal{D}}

\newcommand{\Kcal}{\mathcal{K}}
\newcommand{\Lcal}{\mathcal{L}}

\newcommand{\Ocal}{\mathcal{O}}
\newcommand{\Scal}{\mathcal{S}}

\newcommand{\Ical}{\mathcal{I}}
\newcommand{\Jcal}{\mathcal{J}}

\newcommand{\nfrak}{\mathfrak{n}}

\newcommand{\veps}{\varepsilon}

\newcommand{\zb}{{\bar{z}}}
\newcommand{\fb}{\bar{f}}
\newcommand{\hb}{\bar{h}}

\newcommand{\kcal}{{\mathcal k}}
\newcommand{\ncal}{{\mathcal n}}
\newcommand{\jcal}{{\mathcal j}}

\newcommand{\del}{\partial}
\newcommand{\delbar}{\bar \partial}

\numberwithin{equation}{section}
\begin{document}


 \begin{titlepage}
  \thispagestyle{empty}
 
  \bigskip
  \begin{center}

        \bigskip\ \bigskip\ \bigskip
        \bigskip\ \bigskip\ \bigskip

        \baselineskip=12pt 
        {\LARGE 
           \scshape{Long-Range Interactions in Celestial CFT}
                }
  
        \vskip1.5cm

    \centerline{ 
        Sangmin Choi,
        Ameya Kadhe,
        and Andrea Puhm
    }

\bigskip\bigskip

\centerline{\em {Institute for Theoretical Physics, University of Amsterdam,}}\smallskip
\centerline{\em {PO Box 94485, 1090 GL Amsterdam, The Netherlands}}\bigskip

\smallskip

\bigskip\bigskip
 
\end{center}

    \begin{abstract}
  
Loop corrections in QED and gravity have recently been conjectured to  give rise to an infinite tower of logarithmic soft theorems governing the universal low-energy behavior of photons and gravitons. We explore the implications of this tower for celestial CFT and for the algebra of conformally soft operators. The symmetry-governed part of the tower of logarithmic soft factors is shown to exponentiate, which demonstrates that these loop effects do not represent independent multi-particle interactions, but instead are rooted in the long-range exchange of gauge bosons between pairs of hard operator insertions. We define conformally soft loop operators, and compute their operator product expansions on the celestial sphere. The associated Ward identities exhibit characteristic non-local behaviors, which reflect the pair-wise interactions between hard operator insertions mediated by gauge bosons. We comment on the implications of these results for the soft operator algebra at loop order.

    \end{abstract}

\vfill
\noindent{\em Email: s.choi@uva.nl, a.m.kadhe@uva.nl, a.puhm@uva.nl}
\end{titlepage}

\setcounter{tocdepth}{2}
\tableofcontents

\newpage
\section{Introduction}

Symmetries abound in Nature. They are a most powerful organizing principle in decoding the fundamental laws that govern our world. In four-dimensional asymptotically flat spacetimes which approximates our universe below cosmological scales a particularly rich set of symmetries arises at large distances and late times \cite{Bondi:1962px,Sachs:1962wk,Sachs:1962zza,Barnich:2009se,Barnich:2010ojg}. These asymptotic symmetries explain the universality of soft factorization theorems in quantum field theory and have observable consequences in the form of memory effects.\footnote{See the review \cite{Strominger:2017zoo} and references therein.} 

At tree level the S-matrix element admits a power-law expansion in the soft energy $\omega$. Demanding that the soft factor has manifest SL(2,$\mathbb C$) Lorentz symmetry reveals an infinite tower $\sim \omega^n$ with $n\geq -1$ of soft factors \cite{Guevara:2021abz,Pasterski:2021fjn}  which resum into an exponential \cite{Bautista:2019tdr,Adamo:2021lrv,Himwich:2023njb}. Using conformal representation theory, which makes an appearance because bulk Lorentz symmetry acts as global conformal symmetry on the celestial sphere at the null boundary, the tower of soft operators can be shown to satisfy (higher-order) conservation equations \cite{Pasterski:2021fjn,Pano:2023slc}. These in turn encode the asymptotic symmetry generators of asymptotically flat spacetimes \cite{Donnay:2018neh,Donnay:2020guq}.

In theories with massless particles, the S-matrix is infrared-divergent in four spacetime dimensions due to long-range interactions. For a soft particle attached to the amplitude of $N$ hard matter particles, an infrared-finite observable is obtained by the ratio of the $(N+1)$-particle amplitude over the $N$-particle amplitude \cite{Sahoo-Sen}. The soft expansion becomes non-analytic $\sim \omega^n (\ln \omega)^\ell$ at $\ell$-loop and renders tree-level soft theorems ambiguous. For $n=\ell-1$ a universal tower of logarithmic soft theorems has been conjectured whose soft factors only depend on the momenta of the hard particles but not the details of the theory. These universal towers beg for an asymptotic symmetry interpretation. Beyond the Weinberg soft theorem $\sim \omega^{-1}$ which is tree-exact, the classical limit of the leading logarithmic soft theorem $\sim \ln \omega$ turns out to be associated to superrotations (gravity) and superphaserotations (gauge theory) \cite{Campiglia:2019wxe,Choi:2024ygx,Choi:2024ajz,Choi:2024mac,Choi:2025mzg} which are the same symmetries that govern the subleading tree-level soft graviton and photon theorems \cite{Cachazo:2014fwa,Kapec:2014opa,Lysov:2014csa, Campiglia:2014yka,Campiglia:2015yka,Campiglia:2015kxa,Campiglia:2016hvg}.
For subleading logarithmic soft theorems \cite{Sahoo:2020ryf,Krishna:2023fxg,Karan:2025ndk} the asymptotic symmetry derivation is an open problem.

Logarithmic soft theorems turn out to satisfy the same type of conservation equations as subleading tree-level soft theorems as we show here. Demanding that the tower of logarithmic soft factors has manifest SL(2,$\mathbb C$) symmetry and discarding terms that drop out of the conservation equations (these terms are not encoded in soft symmetry generators), we find that logarithmic soft theorems, too, resum into an exponential. This is reminiscent of the exponentiation of virtual infrared divergences \cite{Weinberg:1965nx} but the form of logarithmically soft exponentiation is different as we discuss.

In theories with gravitational interactions the backreaction of the finite-energy hard particles on the soft radiation creates a gravitational drag \cite{Sahoo-Sen} that shows up as an overall factor in the resummed expression. Its role is fundamentally different from the other contributions to logarithmic soft theorems. While all contributions arise from infrared-regulated loop integrals $\sim \omega^{\ell-1} (\ln \omega)^\ell$, there are two distinct scales involved that render the logarithm dimensionless: the characteristic system size $L$ determined by the hard particle energies, and the largest scale in the scattering process identified as the distance to the detector~$R$. The latter is the scale that enters in the gravitational drag. We interpret the scale $R$ in the context of Faddeev-Kulish dressings \cite{Kulish-Faddeev,Ware:2013zja} which reveals that the energetically soft limit corresponds to expanding $\omega L\ll 1$ around $\omega=1/R$. In the soft limit $\omega R\to 1$ and so the gravitational drag does not contribute. Soft symmetries instead arise from logarithmic contributions $\sim \omega^{\ell-1} (\ln\omega L)^\ell$. One of the goals of this paper is to understand the implications of these symmetry-governed logarithmic soft factors for the holographic dual to quantum gravity in asymptotically flat spacetimes.\footnote{The implications of loop corrections in non-Abelian gauge theories for celestial CFT were recently discussed in \cite{Magnea:2025zut}.}

In quantum gravity the asymptotic symmetries of spacetimes with a negative cosmological constant have been very potent in constraining holographically dual boundary theories \cite{Maldacena:1997re,Witten:1998qj,Aharony:1999ti,Ryu:2006bv}. For zero cosmological constant the set of asymptotic symmetries is much richer, thus offering important constraints on flat space holographic dual theories which are much harder to come by.
The asymptotic symmetries governing the soft factorization of the S-matrix are generated by primary operators with conformally soft dimensions $\Delta \in 1-\mathbb Z_{\geq 0}$ in two-dimensional celestial CFT.
The Ward identities associated to tree-level
soft theorems correspond to insertions of topological charge operators which are constructed from these conformally soft primaries and have a local action on hard operators.
The sum of the variations of each insertion giving
zero defines the symmetry transformation. This is familiar from standard CFT.

As the conjectured holographic dual to quantum gravity in four-dimensional asymptotically flat spacetime \cite{Pasterski:2021raf,Pasterski:2021rjz,Raclariu:2021zjz,McLoughlin:2022ljp}, celestial CFT has to account for long-range interactions. These turn out to render the action of asymptotic symmetries non-local. 
At one loop, the action of the symmetry generator on the $i$-th hard matter operator involves a sum over interactions between pairs $(ij)$ of operators where $j\neq i$ spans all other hard operator insertions. This represents the exchange of gauge bosons between the external legs at long distances. The non-local structure persists to all loop order: at $\ell$ loops, the action on the $i$-th operator involves a sum over $\ell$ pairs $(ij_1),\dots,(ij_\ell)$ of operators, which reflects the long-range interaction between the $i$-th operator and the other hard operators mediated by the gauge boson.
The non-local and state-dependent action of asymptotic symmetries in theories with long-range interactions is not surprising:
charged operators create long-range fields and the charge insertion measures their mutual interactions.
The dependence on all hard operators is what one expects when the charge operator probes the background created by other charged operators.

Despite the non-local action on hard operators, we can nevertheless compute the operator product expansion (OPE) between conformally soft operators on the celestial sphere via consecutive soft limits. In contrast to the tower of soft graviton operators whose singular celestial OPEs were found to generate the $w_{1+\infty}$ symmetry algebra, the singular celestial OPE between universal conformally soft loop operators in gravity turns out to vanish. The same is true for scalar QED, both for tree and loop operators, albeit for a more mundane reason: photons are uncharged.

Beyond the leading soft graviton operator all other conformally soft tree operators receive loop corrections. What are the consequences for the $w_{1+\infty}$ symmetry algebra? To answer this question would likely require the knowledge of an infinite tower of non-universal loop corrections. A more modest target is to understand the loop corrections to the BMS subalgebra of $w_{1+\infty}$ which only involves conformally soft operators with dimension $\Delta=1$ and $\Delta=0$. While the leading $\Delta=1$ soft graviton operator generating supertranslations is tree-exact, the subleading $\Delta=0$ tree soft graviton operator generating superrotations receives corrections at one-loop; these are currently under investigation \cite{Choi-Kadhe-Puhm_wip}. In addition there is a $\Delta=0$ conformally soft loop operator which is one-loop exact. The celestial OPE between this $\Delta=0$ conformally soft loop operator and the $\Delta=1$ and $\Delta=0$ tree operators reveals a subtlety regarding the order of consecutive soft limits that is absent at tree-level \cite{Mitra-Pano-Puhm_toappear}. For a preferred order of soft limits we recover the BMS OPEs where the $\Delta=0$ tree operator is replaced by the universal $\Delta=0$ loop operator.  We leave the study of loop corrections to soft OPEs for the entire tower for future work.\smallskip

This paper is organized as follows. We review
in section~\ref{CCFTsymmetries} how bulk Lorentz symmetry naturally organizes symmetries in celestial CFT using results from conformal representation theory. 
In section~\ref{SoftFactSmatrix} we summarize the soft theorems in scalar QED and gravity for trees and loops, explain what goes logarithmically soft, and show that the symmetry-governed tower of logarithmic soft factors exponentiates for universal loops in a similar way as it does for trees. 
In section~\ref{CelestialWID} we discuss the celestial Ward identities associated with the infinite tower of logarithmic soft photon and graviton theorems.
In section~\ref{SoftOPEs} we compute the celestial OPE between conformally soft operators on the celestial sphere, generalizing the OPEs of the universal tower of tree-level operators to loop level.
We conclude in section~\ref{discussion} with a discussion of our results.
\bigskip

{{\bf Note added:} When this paper was in its final stage, the preprint \cite{Banerjee:2026keq} appeared on the celestial symmetry interpretation of all-loop soft theorems in massless scalar QED which has some overlap with some of our scalar QED results but also differs. While we compute the celestial OPE between conformally soft loop operators associated to logarithmic soft theorems, the authors of \cite{Banerjee:2026keq} focus on the structure of an emergent dipole operator which they argue is necessary for the locality of celestial CFT. It would be interesting to understand how these are connected.\footnote{We thank Shamik Banerjee and Biswajit Sahoo for discussion on this point.}
}

\section{Symmetries in celestial CFT}
\label{CCFTsymmetries}
The universal nature of soft factorization theorems has an asymptotic symmetry origin. These symmetries become visible most clearly in a scattering basis that makes SL(2,$\mathbb C$) Lorentz symmetry manifest. In this section we review how SL(2,$\mathbb{C}$) symmetry serves as a useful organizing principle for celestial CFT that allows to construct conformal primary operators and identify their conservation equations associated to a tower of soft symmetries.

\subsection{SL(2,\texorpdfstring{$\mathbb C$}{C}) symmetry as organizing principle}
\label{SL2C}

The bulk SO(1,3) $\simeq$ SL(2,$\mathbb C$) Lorentz symmetry acts on the celestial sphere at the null boundary as global conformal symmetry
\begin{equation}\label{Sl2Comega}
     z\to f(z)=\frac{az+b}{cz+d},\quad \zb\to \fb(\zb)=\frac{\bar a \zb+\bar b}{\bar c \zb+\bar d},\quad \text{with} \quad  ad-bc=1,\quad \bar a \bar d- \bar b \bar c=1.
 \end{equation}
Making Lorentz symmetry manifest in the scattering data lets us recast the S-matrix as a correlator of conformal primaries which transform under SL(2,$\mathbb C$) as
\begin{equation}\label{OSL2C}
\Ocal_{h,\hb}(z,\zb)\to(\del f)^{-h} (\delbar \fb)^{-\hb}\Ocal_{h,\hb}(z,\zb),
\end{equation}
with weights $(h,\hb)=\Big(\frac{\Delta+J}{2},\frac{\Delta-J}{2}\Big)$. Here $\del \equiv \del_z$ and $\delbar\equiv \del_{\zb}$.

Adopting SL(2,$\mathbb C$) symmetry as a useful organizing principle we parametrize the momenta of massless particles by
\begin{equation}\label{pmassless}
    p^\mu=\eta \omega q^\mu(z,\bar z),\quad q^\mu=\frac{1}{2}\left(1+z\zb, z+\zb, i(\zb-z),1-z\zb\right),
\end{equation}
in terms of their energy $\omega\geq 0$ and a point $(z,\zb)$ on the celestial sphere towards which the particle propagates; the sign $\eta=\pm$ distinguishes between incoming ($-$) and outgoing ($+$) particles.
Under an SL(2,$\mathbb C$) transformation
\begin{equation}\label{Sl2Comegaq}
    \omega \to (\del f)^{-1/2}(\delbar \fb)^{-1/2} \omega,\quad q^\mu(z,\bar z)\to (\del f)^{1/2} (\delbar \fb)^{1/2}\Lambda^\mu_{\;\nu} q^\nu(z,\bar z),
\end{equation}
where $\Lambda$ is the corresponding Lorentz transformation in the vector representation, such that
 \begin{equation}\label{Sl2Cp}
    p^\mu(z,\bar z)\to \Lambda^\mu_{\;\nu} p^\nu(z,\bar z).
\end{equation}
For spin $s$ massless particles we define the polarization vectors for helicity $J=\pm 1$ as 
\begin{equation}\label{veps}
    \veps_+^{\mu}=\sqrt{2}\del q^\mu(z,\zb),\quad \veps_-^{\mu}=\sqrt{2}\delbar q^\mu(z,\zb),
\end{equation}
and construct polarization tensors for helicities $J=\pm 2$ via $\veps^{\mu\nu}_{\pm}=\veps^\mu_{\pm} \veps^\nu_{\pm}$. These polarizations do not transform as SL(2,$\mathbb{C}$) primaries. Instead we define \cite{Himwich:2023njb} 
\begin{equation}\label{vepsprime}
    {\veps'}_{+}^{\mu}=\sqrt{2}\frac{1}{z'-z}q^\mu(z',\zb),\quad  {\veps'}_-^{\mu}=\sqrt{2}\frac{1}{\zb'-\zb}q^\mu(z,\zb'),
\end{equation}
where an auxiliary reference point $(z',\zb')$ was introduced such that under SL(2,$\mathbb{C}$) transformations
\begin{equation}
      {\veps'}_{+}^{\mu}\to (\del f)^{-1/2} (\delbar \fb)^{+1/2}\Lambda^\mu_{\;\nu}  {\veps'}_{+}^{\nu},\quad  {\veps'}_{-}^{\mu}\to (\del f)^{+1/2} (\delbar \fb)^{-1/2}\Lambda^\mu_{\;\nu}  {\veps'}_{-}^{\nu},
\end{equation}
where the reference point is transformed as well.
In the limit $z',\zb'\to\infty$ we recover \eqref{veps}.

Trading the energy $\omega$ for the Lorentz boost weight $\Delta$ via a Mellin transform $\int_0^\infty d\omega \,\omega^{\Delta-1}$
recasts asymptotic energy eigenstates which manifest translation symmetry as asymptotic boost eigenstates which manifest Lorentz symmetry.  The Mellin transform of a plane wave is given by
\begin{equation}\label{CPW}
    \int_0^\infty d\omega\, \omega^{\Delta-1} e^{i\eta\omega q \cdot X-\delta \omega}= \frac{\Gamma(\Delta)}{(-i \eta q\cdot X+\delta)^\Delta}=:\Phi_\Delta^\eta(X;q),
\end{equation}
where $\delta>0$ is a regulator. 
This defines a massless scalar conformal primary wavefunction with conformal dimension $\Delta$. Spinning conformal primary wavefunctions are obtained by dressing the scalar primaries by the appropriate polarizations. 
Dressing \eqref{CPW} with the SL(2,$\mathbb C$)-covariant polarization  vectors ${\veps'}^\mu_\pm$ and polarization tensors ${\veps'}_\pm^{\mu\nu}={\veps'}_\pm^\mu{\veps'}_\pm^\nu$ yields conformal primary wavefunctions $\Phi_{\Delta, J}^\eta(X;z,\zb)$ with SL(2,$\mathbb C$)-spin $s=1$ and $s=2$, respectively. 
Boundary conformal primary operators are obtained via the extrapolate dictionary \cite{Pasterski:2021dqe}
\begin{equation}
\Ocal_{\Delta,J}^\eta(z,\zb):=\Big(O_s(X),\Phi_{\Delta,J}^\eta(X;z,\zb)\Big)_\Sigma
\end{equation}
where $O_s(X)$ is a spin-$s$ bulk quantum field operator and $(.)_\Sigma$ denotes the spin-$s$ generalization of the Klein-Gordon inner product where the integration is over a Cauchy slice $\Sigma$. 
Under SL(2,$\mathbb{C}$) the boundary operator $\Ocal_{\Delta,J}^\eta(z,\zb)$ transforms as a conformal primary operator \eqref{OSL2C}.
Hard matter particles give rise to `conformally hard' primaries with $\Delta \in 1+i \mathbb R$, while soft particles are associated to `conformally soft' primaries with $\Delta \in 1-\mathbb Z_{\geq 0}$.

\subsection{Towers of soft symmetries}
\label{TopoOp}
Operators with conformally soft dimensions $\Delta \in 1-\mathbb Z_{\geq 0}$ generate symmetries in celestial CFT. 
Their conservation equations arise from
primary descendants $O\simeq 0$, where $\simeq$ means equality up to contact terms. There are three types of primary descendants in 2d celestial CFT, depending on whether the spin of the primary descendant is larger ({\rm I}), smaller ({\rm II}) or equal ({\rm III}) in absolute value to that of its parent primary \cite{Pasterski:2021fjn,Pano:2023slc}. The conformal dimensions $\Delta$ for the different types of descendants that become primary are given by
\begin{equation}
\badat{3}
{\rm (I)}:&\quad \Delta=1-s-\nfrak,\quad \nfrak \in \mathbb Z_{>0},\\
{\rm (II)}:&\quad 
\Delta=1+s-\nfrak,\quad \nfrak \in \mathbb \{1,\dots,2s-1\},\\
{\rm (III)}:&\quad\Delta=1-s.
\eadat
\end{equation}
A compact expression that encodes the primary descendant conditions for all types in a unified way is 
\begin{equation}\label{primdesc}
{O_{s-\Delta+1}^\pm\equiv\Dcal^\pm \Ocal_{\Delta,\pm s} \simeq 0,}
\end{equation}
where
\begin{equation}\label{Dcal}
{\Dcal^+= \delbar^{s-\Delta+1},\quad \Dcal^-=\del^{s-\Delta+1}}.
\end{equation}
Insertions of conformally soft primary descendant operators $O_{s-\Delta+1}^\pm$ into correlation functions gives rise to celestial Ward identities which encode the symmetry transformation of a conformally hard operator.

From \eqref{primdesc}-\eqref{Dcal} it follows that we can expand the conformally soft operators with $J=+s$ as 
\begin{equation}\label{Osoftexpansion}
\Ocal_{\Delta,+ s} (z,\zb) =\sum_{m=0}^{s-\Delta} \zb^{s-\Delta-m} \Ocal^{(s-\Delta-m)} (z),
\end{equation}
and define towers of charges $Q^{(m,n)}$ with $n\in \mathbb Z$ and $m=0,\dots, s-\Delta$ as follows
\begin{equation}\label{Qmn}
Q^{(m,n)}=\frac{(-1)^m}{2\pi i} \oint dz\, z^n \Ocal^{(s-\Delta-m)}(z).
\end{equation}
Their insertions into correlation functions deconstructs the conformally soft theorems into correlators with holomorphic behavior. The Noether current
\begin{equation}
\Jcal^\epsilon(z) =\sum_{m=0}^{\ncal-1} (-1)^m \delbar^m \epsilon(z,\zb)\delbar^{s-\Delta-m}\Ocal(z,\zb),
\end{equation}
is conserved, $\delbar \Jcal^\epsilon=0$, if also $\delbar^{s-\Delta+1}\epsilon(z,\zb)=0$. Upon expanding the symmetry parameter as
\begin{equation}
\epsilon(z,\zb)
=\sum_{m=0}^{s-\Delta} \sum_{n\in \mathbb Z} z^n \zb^{m} \epsilon_{m,n},
\end{equation}
the Noether charge can be expressed as 
\begin{equation}
\badat{2}
Q_\Sigma^\epsilon =\frac{1}{2\pi i} \oint_\Sigma dz \Jcal^\epsilon(z)= \sum_{m=0}^{\ncal-1} \sum_{n\in \mathbb Z} m! (s-\Delta-m)! \epsilon_{m,n} Q^{(m,n)}.
\eadat
\end{equation}
A similar analysis for $J=-s$ yields the corresponding charge towers $\bar Q^{(m,n)}$ with antiholomorphic behavior in correlation functions and which give rise to $\bar Q_\Sigma^{\bar \epsilon}=\frac{1}{2\pi i} \oint_\Sigma d\zb \bar \Jcal^{\bar \epsilon}(\zb)$. In the following we review the soft theorems that give rise to this tower of soft symmetries. While at tree-level the charges are topological operators that act locally on the matter, we will see that loops lead to a non-local action.

\section{Soft factorization of the S-matrix}
\label{SoftFactSmatrix}

We begin with a brief summary in section \ref{ReviewSoftTheorems} of the soft factorization theorems of the S-matrix in scalar QED and gravity, contrasting the tree-level power-law expansion with the non-analytic behavior at loop-level due to the appearance of logarithms. At tree-level an SL(2,$\mathbb C$)-covariant contribution to the infinite tower of soft terms has been shown to exponentiate which we review in section~\ref{AllOrderTree}. 
We then identify in section~\ref{AllOrderLoop} a universal contribution to the infinite tower of logarithmic soft theorems that is governed by symmetries and show that it, too, resums into an exponential. We contrast in section~\ref{WhatSoft} the different infrared scales entering in the soft expansion which is important for analyzing the logarithmically soft towers in celestial CFT.

\subsection{Review of soft factors in scalar QED and gravity}\label{ReviewSoftTheorems}

\paragraph{Trees.}
The soft factorization of tree-level scattering amplitudes takes the form of a power-law expansion in the energy~$\omega$,
\begin{equation}\label{SoftTree}
    \mathcal A_{N+1}(p_1,\dots,p_N;p)=\sum_{n=-1}^\infty \omega^n S_n \,\mathcal A_N(p_1,\dots,p_N)+\dots
    ,
\end{equation}
where  $p^\mu=\eta\omega q^\mu$ denotes the soft particle momentum (with $\eta=+1$ for outgoing and $\eta=-1$ for incoming) and $p_i$ denote the momenta of the $i=1,\dots,N$ hard particles. 
The soft factors $S_{n}=\sum_{i=1}^N S_{n,i}$ in gauge theory for $n=-1,0$ are \cite{Weinberg:1965nx, Low:1958sn}
\begin{equation}\label{SoftTreeEM}
\omega^{-1} S^{\rm em}_{-1,i}= e Q_i\frac{\veps\cdot p_i}{p\cdot p_i},\quad  \omega^0S^{\rm em}_{0,i}= eQ_i\frac{\veps \cdot i\Jcal_i\cdot p}{p\cdot p_i},\end{equation}
and in gravity for $n=-1,0,1$ are \cite{Weinberg:1965nx, Cachazo:2014fwa}
\begin{equation}\label{SoftTreeGR}
\omega^{-1}S^{\rm gr}_{-1,i}= \frac{\kappa}{2}\frac{(\veps\cdot p_i)^2}{p\cdot p_i},\quad\omega^0S^{\rm gr}_{0,i}= \frac{\kappa}{2}\frac{(\veps\cdot p_i)(\veps \cdot i\Jcal_i\cdot p)}{p\cdot p_i},\quad  
\omega S^{\rm gr}_{1,i}=\frac{\kappa}{4}\frac{(\veps\cdot i\Jcal_i \cdot p)^2}{p\cdot p_i}.
\end{equation}
Here $\Jcal_i=\Lcal_i+\Sigma_i$ denotes the total angular momentum acting on the $i$-th hard particle which receives contributions from the orbital and spin angular momenta,
\begin{equation}
   \Lcal^{\mu\nu}_i\equiv -i\left(p^\mu_i\partial^\nu_{p_i}-p^\nu_i\partial^\mu_{p_i}\right),\quad  \Sigma^{\mu\nu}_i\equiv -i\left(\veps^\mu_i\partial^\nu_{\veps_i}-\veps^\nu_i\partial^\mu_{\veps_i}\right).
\end{equation}
The soft factors can be shown to be SL(2,$\mathbb C$)-covariant upon using charge, momentum and angular momentum conservation.
Moreover, they are universal in the sense that they depend only on the momenta and angular momenta of the hard particles as well as the direction of the soft particle but not the details of the interactions. 
However, while Weinberg's leading soft photon and graviton theorems have tree-exact soft factors, all subleading soft theorems receive loop-corrections.

\paragraph{Loops.}
The long-range nature of gravitational interactions as well as quantum loop effects give rise to non-analytic behavior in the soft expansion \cite{Laddha:2018myi,Laddha:2018vbn, Sahoo-Sen}
\begin{equation}\label{SoftLoop}
   \mathcal A_{N+1}(p_1,\dots,p_N;p)=\sum_{n=-1}^\infty \omega^n (\ln \omega)^{n+1} S^{\ln}_n\,\mathcal A_N(p_1,\dots,p_N)+\dots,
\end{equation}
where the $\dots$ denote non-universal contributions $\sim \omega^n (\ln \omega)^\ell$ with $\ell\neq n+1$.
Here we focus on universal soft factors $\sim \omega^n (\ln \omega)^{n+1}$ which do not depend on the details of the interactions. For $n=-1$ the soft factor reduces to Weinberg's tree-exact result  $S_{-1}^{\ln}\equiv S_{-1}$. For $n\geq0$ the logarithmic soft factors were shown \cite{Sahoo-Sen} to receive contributions from classical infrared and quantum loop effects,
\begin{equation}\label{quclsplit}
S_n^{\ln}=S_{n,\rm classical}^{\ln}+S_{n,\rm quantum}^{\ln}.
\end{equation}

In (scalar) QED the leading logarithmic soft photon factor is given by \cite{Sahoo-Sen}
\begin{equation}\label{SoftLoopEM0}
\ln \omega \,S^{\ln,\rm em}_{0,i}=e Q_i\frac{(\veps \cdot i\Jcal_i\cdot p)\Kcal_{\rm em}}{p\cdot p_i},
\end{equation}
while in gravity the leading logarithmic soft graviton factor is given by \cite{Sahoo-Sen}
\begin{equation}\label{SoftLoopGR0}
\ln \omega \,S^{\ln,\rm gr}_{0,i}=\frac{\kappa}{2} \Kcal_{\rm drag} \frac{(\veps\cdot p_i)^2}{p\cdot p_i}+\frac{\kappa}{2}\frac{(\veps \cdot p_i)(\veps \cdot i\Jcal_i\cdot p)\Kcal_{\rm gr}}{p\cdot p_i},
\end{equation}
and the subleading logarithmic soft graviton factor is expected to take the form \cite{Krishna:2023fxg} \footnote{This form has been derived in the classical limit \cite{Sahoo:2020ryf}.}
\begin{equation}\label{SoftLoopGR1}
\omega(\ln \omega)^2 \,S^{\ln,\rm gr}_{1,i}=\frac{\kappa}{4} \Kcal_{\rm drag}^2\frac{ (\veps\cdot p_i)^2}{p\cdot p_i}+\frac{\kappa}{2} \Kcal_{\rm drag}\frac{(\veps \cdot p_i)(\veps \cdot i\Jcal_i\cdot p)\Kcal_{\rm gr}}{p\cdot p_i}+\frac{\kappa}{4}  \frac{\left[(\veps \cdot i \Jcal_i \cdot p)\Kcal_{\rm gr}\right]^2}{p\cdot p_i}.
\end{equation}
The soft factors are expressed in terms of
\begin{equation}\label{MassiveKsmomentum}
\badat{3}
  \Kcal_{\rm em} &=\frac{i}{8\pi}e^2 \ln \omega \sum_{i=1}^N \sum_{\substack{j=1\\ j\neq i}}^N  Q_i Q_j\frac{1}{\beta_{ij}}\left[\delta_{\eta_i\eta_j,1}+\frac{i}{2\pi} \ln\left(\frac{1+\beta_{ij}}{1-\beta_{ij}}\right) \right],\\
 \Kcal_{\rm gr}&=\frac{i}{16\pi}\left(\frac{\kappa}{2}\right)^2  \ln \omega \sum_{i=1}^N \sum_{\substack{j=1\\ j\neq i}}^N (p_i\cdot p_j) \frac{1+\beta_{ij}^2}{\beta_{ij}} \left[\delta_{\eta_i\eta_j,1}+\frac{i}{2\pi} \ln\left(\frac{1+\beta_{ij}}{1-\beta_{ij}}\right) \right],
  \eadat
\end{equation}
where we introduced
\begin{equation}
    \beta_{ij}=\sqrt{1-\frac{p_i^2 p_j^2}{(p_i\cdot p_j)^2}},
\end{equation}
and $\eta_i \eta_j =+1$ if both $i$th and $j$th hard particles are either incoming or outgoing.
In gravity, an additional term in the soft factor arises from the gravitational drag, which is given by 
\begin{equation}\label{MassiveKdragmomentum}
 \Kcal_{\rm drag}=  \frac{i}{4\pi}  \left(\frac{\kappa}{2}\right)^2  \ln \omega \sum_{i=1}^N (p\cdot p_i) \left[\delta_{\eta_i,1}-\frac{i}{2\pi} \ln\left(\frac{p_i^2}{(q\cdot p_i)^2}\right)\right],
\end{equation}
where we chose the soft graviton to be outgoing ($\eta=+1$)  and $\eta_j=+1$ if the $j$-th hard particle is outgoing. 
The imaginary (real) terms in \eqref{MassiveKsmomentum} and \eqref{MassiveKdragmomentum} contribute to the classical (quantum) logarithmic soft factor in \eqref{quclsplit}.

\subsection{Trees: universal towers resum}
\label{AllOrderTree}

Beyond subleading order in scalar QED and the sub-subleading order in gravity, a tower of soft theorems was identified in \cite{Hamada:2018vrw,Li:2018gnc} to take the form 
\begin{equation}\label{Sng01}
\badat{2}
 \omega^n S^{\rm em}_{n\geq1,i}&= eQ_i\frac{\veps \cdot i\Lcal_i\cdot p}{p\cdot p_i}\frac{1}{(n+1)!}\left(p\cdot \partial_{p_i}\right)^{n}
,\\  
\omega^n S^{\rm gr}_{n\geq2,i}&=\frac{\kappa}{2}\frac{(\veps\cdot i\Lcal_i \cdot p)^2}{p\cdot p_i}\frac{1}{(n+1)!}\left(p\cdot \partial_{p_i}\right)^{n-1}.
\eadat
    \end{equation}
This tower does not transform as SL(2,$\mathbb{C}$) primaries. A natural proposal \cite{Bautista:2019tdr,Himwich:2023njb} is to work with the SL(2,$\mathbb{C}$)-covariant polarization vectors \eqref{vepsprime} and complete the $p\cdot \del_{p_i}$ factors into angular momentum generators using the Lorentz-covariant polarization vectors,
\begin{equation}\label{completion}
    p\cdot \partial_{p_i}\longrightarrow \frac{1}{\veps'\cdot p_i} \left[(\veps' \cdot p_i)p\cdot \partial_{p_i}-(p\cdot p_i) \veps' \cdot \partial_{p_i}\right] =\frac{\veps' \cdot i \Lcal_i \cdot p}{\veps' \cdot p_i}.
\end{equation}

This yields towers of SL(2,$\mathbb{C}$)-covariant soft factors in scalar QED and gravity which for any $n\geq -1$ can succinctly be expressed as
\begin{equation}\label{SoftTowerPrimary}
  \omega^n {S'}_{n,i}=\omega^{-1} {S'}_{-1,i}\frac{1}{(n+1)!} \left(\omega \frac{{S'}_{0,i}}{{S'}_{-1,i}}\right)^{n+1}.
\end{equation}
Notice that while ${S'}_{-1,i}$ and ${S'}_{0,i}$ differ for soft photons and gravitons, the operator
\begin{equation}
\frac{{S'}_{0,i}}{{S'}_{-1,i}}=\frac{\veps'\cdot i\Jcal_i\cdot q}{\veps'\cdot p_i}
\end{equation}
is the same for both; we replaced $\Lcal_i \to \Jcal_i$ to allow for hard particles with spin.
The leading soft factor ${S'}^\pm_{-1,i}$ for positive or negative helicity soft particles transforms as a primary of weights $(h,\bar h)=\left(\frac{1+J}{2},\frac{1-J}{2} \right)$ , while the ratio $\frac{{S'}^\pm_{0,i}}{{S'}^\pm_{-1,i}}$ transforms as a $(h,\bar h)=\left(-\frac{1}{2},-\frac{1}{2} \right)$ primary. 
The $n$-th tree-level soft factor transforms as 
\begin{equation}
      {S'}^\pm_{n,i}\to (\del f)^{(n-J)/2} (\delbar \fb)^{(n+J)/2} {S'}^\pm_{n,i},
\end{equation}
which is a $(h,\bar h)=\left(\frac{-n+J}{2},\frac{-n-J}{2}\right)$ primary.
The sum of all tree-level soft factors \eqref{SoftTowerPrimary} exponentiates
\begin{equation}
  \sum_{n=-1}^\infty \omega^n {S'}_{n}=\sum_{i=1}^N \frac{1}{\omega} {S'}_{-1,i} \,e^{\omega \frac{{S'}_{0,i}}{{S'}_{-1,i}}}.
\end{equation}
This tower of SL(2,$\mathbb C$)-covariant tree-level soft theorems satisfies the primary descendant condition
\begin{equation}\label{DcalSprime}
\Dcal^\pm {S'}_{n,i}^\pm \simeq 0.
\end{equation}
The relation between the soft factor ${S'}^\pm_{n,i}$ with polarizations \eqref{vepsprime} and ${S}^\pm_{n,i}$ defined as \eqref{SoftTowerPrimary} but with polarizations \eqref{veps} is the factor
\begin{equation}
\label{SprimeoverS}
\frac{{S'}^\pm_{n,i}}{S^\pm_{n,i}}=\left(\frac{{\veps}^\pm \cdot p_i}{{\veps'}^\pm \cdot p_i}\right)^{-s+(n+1)} .
\end{equation}
For $s=1$ the reference point actually drops out of ${S'}^{\rm em}_{-1}$ due to charge conservation, while ${S'}^{\rm em}_{0}$ is independent of $(z',\zb')$. For $s=2$ the reference point also drops out of ${S'}^{\rm gr}_{-1}$ and ${S'}^{\rm gr}_{0}$ by, respectively, momentum and angular momentum conservation, while ${S'}^{\rm gr}_{1}$ is independent of $(z',\zb')$. 
To see that for general $-s+(n+1)\geq0$ the primary descendant is independent of the reference point notice that in the parametrization \eqref{pmassless} we have
\begin{equation}\label{eps0overeps}
\frac{\veps^+\cdot p_i}{{\veps'}^+\cdot p_i}=  \frac{z-z'}{z_i-z'},\quad \frac{\veps^-\cdot p_i}{{\veps'}^-\cdot p_i}=  \frac{\zb-\zb'}{\zb_i-\zb'}.
\end{equation}
The factor $(\frac{{\veps}^\pm \cdot p_i}{{\veps'}^\pm \cdot p_i})^{-s+(n+1)}$ for positive (negative) helicity thus vanishes for $-s+(n+1)\geq0$ when hit by $\delbar$ ($\del$), and simplifies to unity when evaluated at the coincident point $(z_i,\zb_i)$ in the contact term in~\eqref{DcalSprime}. Notice also that the completion term in \eqref{completion} $\sim p\cdot p_i$ cancels the pole $\sim \frac{1}{p\cdot p_i}$ in the soft factor and so cannot contribute to the residue entering the Ward identity. 
This shows that the tower of tree-level soft theorems satisfies 
\begin{equation}
\Dcal^\pm S_{n,i}^\pm \simeq 0,
\end{equation}
whose insertion in correlators yields Ward identities.
The asymptotic symmetries associated to these soft theorems are for $s=1$ given by superphaserotations ($n=-1$) \cite{Kapec:2015ena,He:2014cra}, and a tower of divergent superphaserotations ($n\geq 0$) \cite{Campiglia:2016hvg}, while for $s=2$ we have supertranslations ($n=-1$) \cite{Strominger:2013jfa,He:2014laa}, superrotations ($n=0$) \cite{Campiglia:2014yka}, and a tower of divergent superrotations ($n\geq 1)$ \cite{Campiglia:2016efb,Balderston-Choi-Puhm}. 

The benefit of the tower~\eqref{SoftTowerPrimary} is that the soft factors transforms as SL(2,$\mathbb{C}$) primaries. 
In summary, defining $\Scal_n=\omega^n S'_n$, we can write the exponentiated tree-level soft tower compactly as
\begin{equation}\label{ExpSoftTree}
{  \sum_{n=-1}^\infty  {\Scal}_{n}=\sum_{i=1}^N  \Scal_{-1,i} \,e^{\frac{\Scal_{0,i}}{\Scal_{-1,i}}}}.
\end{equation}
The form \eqref{ExpSoftTree} is convenient for attempting to extract a similar structure for all universal loops.

\subsection{Loops: universal towers resum too}
\label{AllOrderLoop}

The tower of logarithmic soft factors $\sim \omega^n (\ln \omega)^{n+1}$ is conjectured to be universal but their explicit form, beyond the cases\footnote{A non-universal term of order $O(\omega \ln \omega)$ at 1-loop which receives corrections at 2-loop has also been conjectured in~\cite{Krishna:2023fxg}. We thank Biswajit Sahoo for pointing out that its classical limit had been derived earlier \cite{Ghosh:2021bam}. } discussed in section~\ref{ReviewSoftTheorems}, is not known in general. Here we identify a universal SL(2,$\mathbb{C}$)-covariant tower of logarithmic soft factors in scalar QED and gravity that is governed by symmetries and exponentiates. This is based on the the known structure of the tower of classical logarithmic soft photon theorems, and we assume that the quantum contribution alters it only through the extra real contribution inside $\Kcal_{\rm em}$ as in \eqref{MassiveKdragmomentum}. In gravity we assume the structure proposed in \cite{2008.04376} for the classical soft graviton tower; our results continue to hold in the quantum theory if the quantum logarithmic soft graviton tower only adds an extra real contribution inside $\Kcal_{\rm gr}$ as in \eqref{MassiveKdragmomentum}.
Our analysis applies to both massive and massless matter.

\paragraph{Soft photon towers.}

In equation (5.9) of \cite{Sahoo:2020ryf} an expression for the $n$-th classical soft photon factor $\sim \omega^n (\ln \omega)^{n+1}$ was proposed, which can succinctly be expressed as
\begin{equation}
    \Scal^{\ln,\rm em}_{-1}=\sum_{i=1}^N \Scal_{-1,i}^{\rm em}\,,\quad
    \Scal^{\ln,\rm em}_{0}=\sum_{i=1}^N \Scal_{0,i}^{\rm em}\Kcal_{\rm em},
\end{equation}
for $n=-1,0$ while for $n\geq1$ the all-loop expression is
\begin{equation}\label{SoftTowerBEM}
\badat{3}
          \Scal^{\ln,\rm em}_{n\geq 1}&=\sum_{i=1}^N \Scal_{0,i}^{\rm em}\Kcal_{\rm em} \frac{1}{(n+1)!}\left[p\cdot \del_{p_i} \Kcal_{\rm em}\right]^{n} \\
         &\quad+\sum_{i=1}^N \omega^n[\ln (\omega+i\epsilon \eta_i)]^{n+1} q_{\alpha_1}\cdots q_{\alpha_{n}}\veps_\mu  \Bcal^{(n+1),\mu\alpha_1\cdots \alpha_{n}} .
\eadat
\end{equation}
Here $\Bcal^{(n+1),\mu\alpha_1\cdots \alpha_{n}} (Q_i,p_i)$ is antisymmetric under $\mu\leftrightarrow \alpha_i$ exchange for $i=1,\dots,n$ and is expected to be independent of the details of the scattering process. Due to the polynomial structure in $q\cdot p_i$ of the second line in \eqref{SoftTowerBEM}, terms involving $\Bcal^{(n+1),\mu\alpha_1\cdots \alpha_{n}}$ do not contribute to the soft photon Ward identity (at any loop order) given by the contact terms in $\Dcal^\pm {S}^{\rm \ln,em}_{n,i}\simeq 0$. Thus for the purpose of extracting the universal tower governed by symmetries we can drop their contribution.

The remaining tower given in the first line of \eqref{SoftTowerBEM} is of a similar form as \eqref{Sng01} and it is natural to consider the completion~\eqref{completion}. 
This leads to a universal tower of SL(2,$\mathbb C$)-covariant logarithmic soft photon factors where the $n$-th soft factor for any $n\geq -1$ is given by
\begin{equation}
    \Scal^{\ln,\rm em}_{n}=\sum_{i=1}^N \Scal_{-1,i}^{\rm em} \frac{1}{(n+1)!} \Big[\frac{\Scal_{0,i}}{\Scal_{-1,i}} \Kcal_{\rm em}\Big]^{n+1}.
\end{equation}
The sum over all universal logarithmic soft factors exponentiates
\begin{equation}\label{EMExpUniversalTower}
     \boxed{\sum_{n=-1}^\infty \Scal^{\ln,\rm em}_{n}=\sum_{i=1}^N \Scal^{\rm em}_{-1,i} \,e^{\frac{\Scal_{0,i}}{\Scal_{-1,i}} \Kcal_{\rm em}}}.
\end{equation}
The all-order universal logarithmic soft photon theorem encodes a remarkably simple symmetry-governed structure! 
The classical limit of these logarithmic soft theorems can be related to asymptotic symmetries.
For $n=-1$ we have $\Scal_n^{\rm \ln,em}\equiv \Scal_n^{\rm em}$ which is associated to superphaserotations \cite{Choi:2024ygx,Choi:2024mac,Kapec:2015ena,He:2014cra}, while for $n=0$ the  symmetry associated to $\Scal_0^{\rm \ln, em}$ is divergent superphaserotations \cite{Choi:2024mac} --- the same asymptotic symmetry as for the subleading tree-level soft photon theorem with $\Scal^{\rm em}_0$ \cite{Campiglia:2016hvg}; it is natural to expect that the tower of (ever more diverging) asymptotic symmetries associated to tree-level soft theorems with $\Scal^{\rm em}_{n>0}$ is also the tower of symmetry underlying the logarithmic soft theorems with $\Scal^{\rm \ln,em}_{n>0}$.

\paragraph{Soft graviton towers.}

In equation (5.13) of \cite{Sahoo:2020ryf}, an expression for the $n$-th classical soft graviton factor $\sim \omega^n (\ln \omega)^{n+1}$ was proposed.
After expressing the leading  $n=-1,0,1$ soft factors in the form
\begin{equation}
      \Scal^{\ln, \rm gr}_{-1}=\sum_{i=1}^N \Scal_{-1,i}^{\rm gr},\;
    \Scal^{\ln, \rm gr}_{0}=\sum_{i=1}^N \Scal^{\rm gr}_{-1,i}\left[\frac{\Scal_{0,i}}{\Scal_{-1,i}}\Kcal_{\rm gr}+\Kcal_{\rm drag}\right],\; 
    \Scal^{\ln, \rm gr}_{1}=\sum_{i=1}^N \Scal^{\rm gr}_{-1,i}\frac{1}{2}\left[\frac{\Scal_{0,i}}{\Scal_{-1,i}}\Kcal_{\rm gr}+\Kcal_{\rm drag}\right]^2,
\end{equation}
 the suggested all-loop expression for $n\geq2$ can be massaged into a compact and suggestive form 
\begin{equation}\label{SoftTowerCGR}
\badat{2}
 \Scal^{\ln,\rm gr}_{n\geq2}&=\sum_{i=1}^{N} \Big\{\frac{1}{(n+1)!}\left[\Kcal_{\rm drag}\right]^{n+1}\Scal^{\rm gr}_{-1,i}+\frac{1}{n!}\left[\Kcal_{\rm drag}\right]^n\Scal^{\rm gr}_{-1,i}\left[\frac{\Scal_{0,i}}{\Scal_{-1,i}}\Kcal_{\rm gr}\right]\\
 &\quad\quad+\sum_{l=2}^{n+1}\frac{1}{(n+1-l)!}\left[\Kcal_{\rm drag}\right]^{n+1-l}\Scal^{\rm gr}_{-1,i}\left[\frac{\Scal_{0,i}}{\Scal_{-1,i}}\Kcal_{\rm gr}\right]^2\frac{1}{l!} \Big[p\cdot \del_{p_i}\Kcal_{\rm gr}\Big]^{l-2}+\\
  &\quad\quad+  \sum_{l=3}^{n+1}  \frac{1}{(n+1-l)!} \left[\Kcal_{\rm drag}\right]^{n+1-l}\omega^{n} \left[\ln (\omega+i\epsilon \eta_i)\right]^l q_{\alpha_1} \cdots q_{\alpha_{l-1}}\veps_\mu \veps_\nu \Ccal^{(l),\mu\nu\alpha_1\cdots \alpha_{l-1}}\Big\}.
      \eadat
      \end{equation}
Here $\Ccal^{(l),\mu\nu\alpha_1\cdots \alpha_{l-1}}(p_i)$ is anti-symmetric under the exchange $\mu\leftrightarrow \alpha_i$ as well as $\nu\leftrightarrow \alpha_i$ for $i,j=1,\dots,n$ and is expected to be be independent of the details of the scattering. 
Due to the polynomial structure in $q\cdot p_i$ of the third line in \eqref{SoftTowerCGR}, terms involving  $\Ccal^{(l),\mu\nu\alpha_1\cdots \alpha_{l-1}}$ do not contribute to the soft graviton Ward identity (at any loop order) given by the contact terms in $\Dcal^\pm {S}^{\rm \ln,gr}_{n,i}\simeq 0$. For the purpose of extracting the universal tower governed by symmetries we can drop their contribution.

The remaining tower given by the first two lines of \eqref{SoftTowerCGR} takes a similar form to \eqref{Sng01} and it is again natural to consider the completion~\eqref{completion}. This leads to a  tower of SL(2,$\mathbb C$)-covariant logarithmic soft graviton factors where the $n$-th universal soft factor for any $n\geq -1$ is given by
\begin{equation}
\badat{2}
    \Scal^{\ln,\rm gr}_{n}    &=\sum_{i=1}^N \Scal^{\rm gr}_{-1,i}\frac{1}{(n+1)!}\left[\frac{\Scal_{0,i}}{\Scal_{-1,i}}\Kcal_{\rm gr}+\Kcal_{\rm drag}\right]^{n+1}.
   \eadat
\end{equation}
The sum over all such universal soft factors exponentiates\footnote{Incidentally, a version of this formula for the case of $2\to 2$ scattering  in the classical limit has been found in \cite{Alessio:2024onn}. We thank Carlo Heissenberg for pointing this out.}
\begin{equation}\label{GRExpUniversalTower}
   \boxed{ \sum_{n=-1}^\infty  \Scal^{\ln,\rm gr}_{n}=e^{\Kcal_{\rm drag}}\sum_{i=1}^N \Scal^{\rm gr}_{-1,i} e^{\frac{\Scal_{0,i}}{\Scal_{-1,i}}\Kcal_{\rm gr}}}.
\end{equation}
The structure of the symmetry-governed tower of the all-order universal logarithmic soft graviton theorem is again strikingly simple! The novelty compared to gauge theory is the gravitational drag contribution which, however, appears only as an overall factor. The classical limit of these logarithmic soft theorems can again be related to asymptotic symmetries. 
For $n=-1$ we have $\Scal^{\rm \ln,gr}\equiv \Scal^{\rm gr}_n$ which is associated to supertranslations \cite{Choi:2024ygx,Strominger:2013jfa,He:2014laa}, for $n=0$ the symmetry associated to $\Scal_0^{\rm \ln,gr}$ are superrotations \cite{Choi:2024ajz} --- the same asymptotic symmetry as for the subleading tree-level soft graviton theorem with $\Scal_0^{\rm gr}$ \cite{Cachazo:2014fwa,Kapec:2014opa}; it is natural to expect that the tower of (ever more) divergent asymptotic symmetries associated to tree-level soft theorems with $\Scal^{\rm gr}_{n>0}$ also governs the logarithmic soft graviton theorems with $\Scal_{n>0}^{\rm \ln,gr}$.

\paragraph{Einstein-Maxwell.}

If we include gravitational interactions, the $n$-th all-loop soft photon factor \eqref{SoftTowerBEM} with $n\geq 1$ gets modified by gravitational drag terms similar to \eqref{SoftTowerCGR} and we should replace $\Kcal_{\rm em}\to\Kcal_{\rm em}+\Kcal_{\rm gr}$ .
This leads to an SL(2,$\mathbb C$)-covariant soft photon tower whose sum over all $n$ again exponentiates
\begin{equation}
\sum_{n=-1}^\infty \Scal^{\ln,\rm em}_{n}=e^{\Kcal_{\rm drag}}\sum_{i=1}^N \Scal^{\rm em}_{-1,i} \,e^{\frac{\Scal_{0,i}}{\Scal_{-1,i}} (\Kcal_{\rm em}+\Kcal_{\rm gr})}.
\end{equation}
Similarly, including electromagnetic interactions, the $n$-th all-loop soft graviton factor \eqref{SoftTowerCGR} with $n\geq 2$ is modified by replacing $\Kcal_{\rm gr}\to\Kcal_{\rm gr}+\Kcal_{\rm em}$ .
This leads to an SL(2,$\mathbb C$)-covariant soft graviton tower whose sum over all $n$ again exponentiates
\begin{equation}
\sum_{n=-1}^\infty \Scal^{\ln,\rm gr}_{n}=e^{\Kcal_{\rm drag}}\sum_{i=1}^N \Scal^{\rm gr}_{-1,i} \,e^{\frac{\Scal_{0,i}}{\Scal_{-1,i}} (\Kcal_{\rm gr}+\Kcal_{\rm em})}.
\end{equation}
Recall that the operator $\frac{\Scal_{0,i}}{\Scal_{-1,i}}$ is the same for both soft photons and gravitons, 
and so the only difference between the symmetry-governed tower logarithmic soft photon and graviton towers in theories with electromagnetic and gravitational interactions is the Weinberg soft factor! The gravitational drag common to both towers and is a reflection of the equivalence principle.

\smallskip
The tower of logarithmic soft theorems $\sim\omega^n (\ln \omega)^{n+1}$ thus exhibit a beautifully simple structure governed by symmetry.
Our next goal is to determine the consequences of this exponentiated structure for celestial CFT. But before doing so we must discuss infrared scales and the different roles of the infrared-regulated loop integrals entering $\Kcal_{\rm em}$, $\Kcal_{\rm gr}$ and $\Kcal_{\rm drag}$ in the soft limit.

\begin{figure}
    \centering
    \begin{subfigure}{0.23\textwidth}
        \centering
        \includegraphics[width=\linewidth]{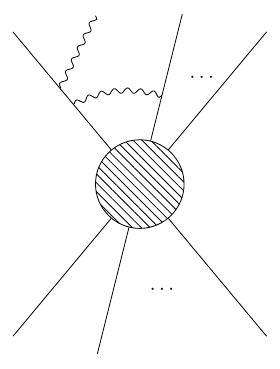}
        \caption{}
        \label{fig:non-drag}
    \end{subfigure}
    \hspace{.1\textwidth}
    \begin{subfigure}{0.23\textwidth}
        \centering
        \includegraphics[width=\linewidth]{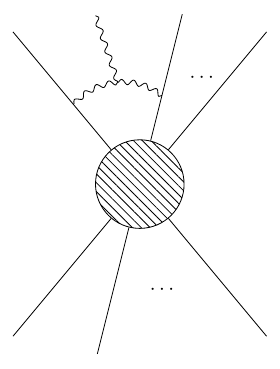}
        \caption{}
        \label{fig:drag}
    \end{subfigure}
    \caption{
        Examples of one-loop diagrams with multiple external scalars and one soft graviton, that give rise to terms logarithmic in the soft graviton energy. Straight lines represent scalars and wiggly lines represent gravitons. (a) This diagram contributes a logarithm in the region where the loop momentum lies between the soft energy $\omega$ and the characteristic energy scale $1/L$ of the (hard) scalars. The soft graviton is emitted from a scalar leg. (b) This type of diagram contributes a logarithm in the region where the loop momentum is less than the soft energy $\omega$.
        The soft graviton is emitted from a graviton loop, so these terms can be interpreted as describing the propagation of a soft graviton through a spacetime that has been backreacted by the external scalars.
    }
    \label{fig:dragnondrag}
\end{figure}

\subsection{What goes logarithmically soft?}
\label{WhatSoft}

The statement that in the soft limit $\omega$ is taken small is by itself not meaningful because $\omega$ is not dimensionless; we must specify a reference energy scale with respect to which $\omega$ is small. There are two relevant physical infrared scales \cite{Sahoo-Sen}: the characteristic system size $L$ and the distance scale to the detector $R$ where $\frac{L}{R}\ll 1$. The loop integrals contributing to the soft expansion split schematically as
\begin{equation}
\Ical^L\sim \int_{\omega}^{1/L} \frac{d\ell}{\ell}\sim \ln(\omega L)\,,\quad\quad \text{and}\quad \quad \Ical^R\sim \int_{1/R}^\omega \frac{d\ell}{\ell} \sim \ln(\omega R)\,.
\end{equation}
The soft photon and graviton contributions  \eqref{MassiveKsmomentum} to the logarithmic soft factors,
\begin{equation}
\badat{2}
   \Kcal_{\rm em} &= \frac{i}{2}e^2\sum_{i=1}^N \sum_{\substack{j=1\\ j\neq i}}^N Q_i Q_j (p_i \cdot p_j) \Ical^L(p_i,p_j),\\
    \Kcal_{\rm gr} &= \frac{i}{2}\left(\frac{\kappa}{2}\right)^2 \sum_{i=1}^N \sum_{\substack{j=1\\ j\neq i}}^N (p_i\cdot p_j)^2 \frac{(1+\beta_{ij}^2)}{2}\Ical^L(p_i,p_j),
\eadat
\end{equation}
have their origin in the infrared-regulated loop integral \cite{Sahoo-Sen}
\begin{equation}\label{Iij}
\badat{2}
   \Ical^L(p_i,p_j)&=\int_\omega^{1/L} \frac{d^4\ell}{(2\pi)^4} \frac{1}{\ell^2-i\epsilon}\frac{1}{(p_i\cdot \ell+i\epsilon)(p_j\cdot \ell -i\epsilon)}\\
    &=\ln( \omega L)\frac{1}{4\pi} \frac{1}{(p_i\cdot p_j)\beta_{ij}} \left[\delta_{\eta_i\eta_j,1}+\frac{i}{2\pi} \ln\left(\frac{1+\beta_{ij}}{1-\beta_{ij}}\right) \right].
\eadat
\end{equation}
The upper bound of integration, $1/L$, represents the order of the energy of the hard particles, or $L$ the system size.
An example of a diagram that contribute to $\Kcal_{\rm gr}$ is given in figure \ref{fig:non-drag}; there are analogous diagrams involving photons that contributes to $\Kcal_{\rm em}$.
A key feature of diagrams that contribute to $\Kcal_{\rm gr}$ and $\Kcal_{\rm em}$ is that the soft particle is emitted from an external leg, which makes it clear that these terms reflect the Coulombic interactions between the soft and hard particles.

In contrast, the contribution \eqref{MassiveKdragmomentum} from the gravitational drag,
\begin{equation}
   \Kcal_{\rm drag} = i\left(\frac{\kappa}{2}\right)^2 \sum_{j=1}^N (p\cdot p_j)^2 \Ical^R(p,p_j), 
\end{equation}
has its origin in the infrared-regulated loop integral \cite{Sahoo-Sen}
\begin{equation}\label{Ii}
\badat{2}
    \Ical^R(p,p_{j})&=\int_{1/R}^\omega \frac{d^4\ell}{(2\pi)^4} \frac{1}{\ell^2-i\epsilon}\frac{1}{(p\cdot \ell+i\epsilon)(p_j\cdot \ell -i\epsilon)}\\
    &=\ln (\omega R)\,\frac{1}{4\pi} \frac{1}{(p\cdot p_j)} \left[\delta_{\eta_j,1} - \frac{i}{2\pi} \ln\left(\frac{p_j^2}{(q\cdot p_j)^2}\right)\right].
    \eadat
\end{equation}
An example of a diagram that contributes to $\Kcal_{\rm drag}$ is given in figure \ref{fig:drag}.
A key feature of such diagrams is that the soft particle interacts with the loop graviton.
Thus, $\Kcal_{\rm drag}$ represents the propagation of the soft particle through spacetime that has been backreacted by the external particles.

\begin{figure}
    \centering
    \begin{subfigure}{.22\linewidth}
        \centering
        \includegraphics[width=\linewidth]{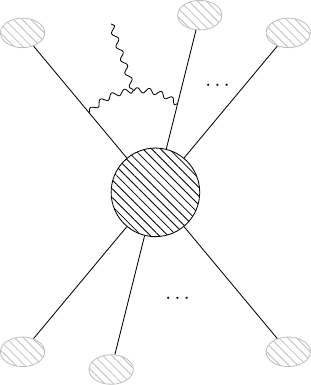}
        \caption{}
        \label{fig:dd1}
    \end{subfigure}
    \hfill
    \begin{subfigure}{.22\linewidth}
        \centering
        \includegraphics[width=\linewidth]{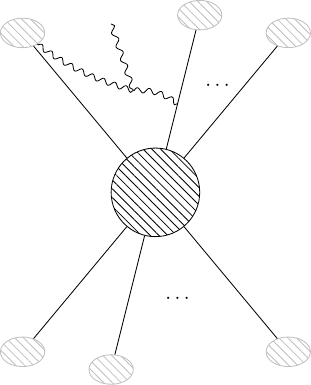}
        \caption{}
        \label{fig:dd2}
    \end{subfigure}
    \hfill
    \begin{subfigure}{.22\linewidth}
        \centering
        \includegraphics[width=\linewidth]{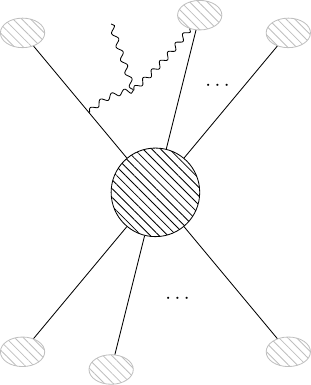}
        \caption{}
        \label{fig:dd3}
    \end{subfigure}
    \hfill
    \begin{subfigure}{.22\linewidth}
        \centering
        \includegraphics[width=\linewidth]{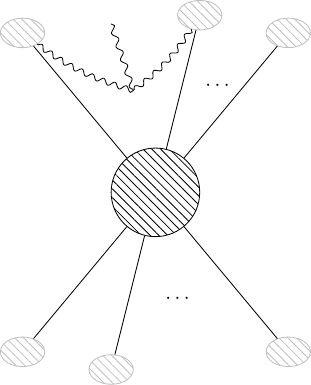}
        \caption{}
        \label{fig:dd4}
    \end{subfigure}
    \caption{One-loop diagrams that contribute logarithms in the presence of graviton dressings. Straight and wiggly lines represent scalars and gravitons respectively, and the gray ellipse at the end of each external leg represents the gravitational dressing of that leg. Because the dressings carry gravitons of energy less than the energy scale $1/R$ associated with the detector distance $R$, the domain of these loop integrals in figures (b)--(d) are restricted to momentum below the scale $1/R$. The diagram in (a) contributes the same term as the one in figure \ref{fig:drag}, while the sum of diagrams (b)--(d) removes the region of loop integral in this contribution below energy scale $1/R$.}
    \label{fig:dressings}
\end{figure}

The lower bound of integration, $1/R$, is interpreted by the authors of \cite{Sahoo-Sen} to be the IR cutoff set by the detector resolution.
This interpretation of the IR regulator serving as the energy scale associated with detector distance is reasonable in the context of Faddeev-Kulish dressed states \cite{Kulish-Faddeev, Ware:2013zja}.
We present an outline of the argument here, and leave a detailed analysis for future work.

In the dressed asymptotic Hilbert space, each particle is accompanied by a cloud of soft gravitons whose energy is below the characteristic time scale $1/T$ of the scattering experiment, which is naturally of the same order as the scale associated with the distance to the detector $1/R$.
The effect of such clouds is that it screens any Weinberg soft poles from real graviton radiation of energy below the scale $1/R$, and cancels contributions from poles in the graviton loops below the energy scale $1/R$ (see \cite{Chung:1965zza, Ware:2013zja, Choi:2017bna}).

One can see the cancellation of pole contributions in loops by considering the diagrams in figure \ref{fig:dressings} that arise when working with the dressed Hilbert space.
The diagram in figure \ref{fig:dd1} corresponds to the one in figure \ref{fig:drag} in the standard Fock state basis; it contributes a logarithm when the loop momentum is less than the soft energy.
The diagrams in figures \ref{fig:dd2}--\ref{fig:dd4} are new; the sum of these three diagrams cancels the lower region of the contribution from \ref{fig:dd1} below energy $1/R$.
Thus, the net effect of introducing gravitational dressings is to replace the lower bound of the loops in the diagram \ref{fig:drag} from the IR cutoff to $1/R$.

It is therefore anticipated that, as far as pole contributions in graviton loops are concerned, the net effect of Faddeev-Kulish dressings is to replace the unphysical IR cutoff (which eventually has to be removed) with the energy scale $1/R$ associated with the detector distance.
Since the dressings screen poles from soft emissions below energy scale $1/R$, the soft expansion should be interpreted as the expansion in the small parameter $\omega L\ll 1$ around $\omega = 1/R$. In this soft limit the combination $\omega R$ never becomes too small and instead $\omega R\to 1$. 
The arguments above support our hypothesis that the actual 
logarithmically soft term in the soft expansion comes only from $\ln(\omega L)$ and not from $\ln(\omega R)$, and that the gravitational drag term
\begin{equation}
    e^{\Kcal_{\rm drag}}=e^{\left(\frac{\kappa}{2}\right)^2{\omega\ln(\omega R)2\gamma_{\rm gr}\sum_j (q\cdot p_j)\left[ i\pi \delta_{\eta_j,1}-\ln \left(q\cdot p_j\right)\right] }}
\end{equation}
is on a fundamentally different footing compared to the other logarithmic soft factors.

\section{Celestial Ward identities}
\label{CelestialWID}

In this section we discuss the celestial Ward identities associated to logarithmic soft theorems in scalar QED and gravity coupled to massless matter. 
Trading the energy $\omega$ for the Lorentz boost weight $\Delta$ via a Mellin transform $\int_0^\infty d\omega \,\omega^{\Delta-1}$ for each external particle recasts the momentum-space S-matrix as a celestial amplitude
\begin{equation}
\langle \Ocal_{\Delta_1,J_1} \cdots \Ocal_{\Delta_N,J_N}\rangle = \prod_{i=1}^N \int_0^\infty d\omega_i\, \omega_i^{\Delta_i-1}  \Acal(p_1,\cdots, p_N;J_1,\dots, J_N).
\end{equation}
In this basis the S-matrix transforms under SL(2,$\mathbb C$) as
\begin{equation}
\langle \Ocal_{\Delta_1,J_1} \cdots \Ocal_{\Delta_N,J_N}\rangle\to \Big[\prod_{i=1}^N (\del f)^{-h_i}(\delbar \fb)^{-\hb_i}\Big] \langle \Ocal_{\Delta_1,J_1} \cdots \Ocal_{\Delta_N,J_N}\rangle.
\end{equation}
We will abbreviate by $\langle \prod_{i=1}^N \Ocal_{\Delta_i,J_i}\rangle\equiv\langle\dots \rangle$ the celestial amplitude of conformally hard operators $\Ocal_{\Delta_i,J_i}(z_i,\zb_i)$ with $\Delta_i \in 1+i\mathbb R$.
The insertion of an energetically soft particle in the momentum-space amplitude maps to the insertion of a conformally soft operator $\Ocal_{\Delta,J}(z,\zb)$ with $\Delta\in 1-\mathbb Z_{\geq 0}$ in the celestial amplitude and we will write $\langle \Ocal_{\Delta,J} \prod_{i=1}^N \Ocal_{\Delta_i,J_i}\rangle\equiv\langle \Ocal_{\Delta,J}\dots \rangle$.

Celestial amplitudes with conformally soft tree operators were shown to yield celestial Ward identities associated to a tower of classical soft symmetries. In this section we define conformally soft loop operators associated with logarithmic soft theorems, compute their celestial OPEs with conformally hard particles and discuss the associated celestial symmetries.
Our focus will be on positive helicity soft photons and gravitons, but analogous results are obtained for negative helicity.

\subsection{Conformally soft tree and loop operators}
\label{SoftTreeLoopOps}
To recast energetically soft theorems as celestial Ward identities, it will be convenient to represent the momentum-conserving Dirac-delta as
\begin{equation}
    (2\pi)^4 \delta^{(4)}\Big(\sum_{i=1}^N p_i+p\Big)=\int d^4X e^{i\eta \omega q\cdot X} \prod_{i=1}^N e^{i\eta_i \omega_i q_i \cdot X}.
\end{equation}
In the soft limit
we have $\lim_{\omega \to 0} e^{i\eta \omega q\cdot X}=1+i\eta \omega q\cdot X+O(\omega^2)$.
Only the leading $O(
\omega^0)$ term enters the tower of universal logarithmic soft theorems $\sim \omega^n (\ln \omega)^{n+1}$, while higher orders $O(\omega^{>0})$ only affect the subleading power-law soft theorems as well as non-universal subleading logarithmic soft theorems.
In the following we consider outgoing soft particles ($\eta=+1$) and omit the $\eta$ label.

\subsubsection{Conformally soft tree operators}

Upon the Mellin transform, factors of powers of energy $\omega^n$ from the soft expansion correspond to shifts in the conformal dimension as
\begin{equation}\label{Mellinpower}
    \int_0^\infty d\omega\, \omega^{\Delta-1} e^{i\omega q \cdot X}\omega^n= \Phi_{\Delta+n}(X;q).
\end{equation}
This can be recast as the action of the momentum operator expressed in the celestial basis $P^\mu=q^\mu e^{\partial_\Delta}$ so that $ \Phi_{\Delta+n}= e^{n\del_\Delta}\Phi_{\Delta}$.
The tree-level soft theorems $\sim \omega^n$ get mapped to {\it simple poles} at 
\begin{equation}
\Delta= -n\in 1-\mathbb Z_{\geq 0}
\end{equation} 
in the celestial amplitude as can be seen from 
\begin{equation}
\lim_{\Delta\to -n}\frac{1}{(n+1)!}\Phi_{\Delta+n}=
    \frac{(-1)^{n+1}}{\Delta+n} +O(1).
\end{equation}
This corresponds to the insertion of the conformally soft primary operator 
\begin{equation}\label{TreeOsoft}
 \boxed{\Ocal_{-n,\pm s}^{\rm tree}:=\lim_{\Delta\to -n} (-1)^{n+1}(\Delta+n) \Ocal_{\Delta,\pm s}\Big|_{g_s}}
\end{equation}
in the celestial amplitude where $g_{s}$ denotes the coupling ($e$ for $s=1$ and $\frac{\kappa}{2}$ for $s=2$).
From the tree-level soft theorems it can be seen that the operators \eqref{TreeOsoft} obey a conservation equation $\Dcal^\pm \Ocal_{-n,\pm s}^{\rm tree}\simeq 0$ associated to the tower of soft symmetries discussed in section \ref{AllOrderTree}.

\subsubsection{Conformally soft loop operators}
Logarithms in $\omega$, of which there is one for every loop $\ell$, can be encoded in derivatives in $\Delta$,
\begin{equation}\label{Mellinlog}
\int_0^\infty d\omega\, \omega^{\Delta-1} e^{i\eta\omega q \cdot X} \omega^n (\ln\omega)^{\ell}=
   \partial_\Delta^{\ell} \Phi_{\Delta+n}^\eta(X;q),
\end{equation}
Universal logarithmic towers arise for $n=\ell-1$ and are the focus here.
Logarithmic soft theorems show up as {\it higher-order poles}. Upon taking the conformally soft limit 
\begin{equation}
\Delta= -n=1-\ell\in 1-\mathbb Z_{\geq 0},
\end{equation}
the universal tower
$\omega^{\ell-1} (\ln \omega)^{\ell}$ with $\ell\geq 0$ gets mapped to poles of order 
$\ell+1$ as can be seen from
\begin{equation}
\lim_{\Delta\to 1-\ell}\partial_\Delta^{\ell}\Phi_{\Delta+1-\ell}^\eta=
    \frac{(-1)^{\ell}\ell!}{(\Delta-1+\ell)^{\ell+1}} +O(1).
\end{equation}
For $\ell=0$ this coincides with the tree-level simple pole at $n=-1$, while for $\ell>0$ the order of the pole at $n\geq 0$ increases with each loop order $\ell$.
In the celestial amplitude this corresponds to the insertion of the $\ell$-loop conformally soft primary operator 
\begin{equation}\label{LoopOsoft}
  \boxed{\Ocal^{\ell-{\rm loop}}_{1-\ell,\pm s}:=   \lim_{\Delta\to 1-\ell} (\Delta-1+\ell)^{\ell+1} \frac{(-1)^\ell}{\ell!} \Ocal_{\Delta,\pm s}\Big|_{g_s^{2\ell+1}}},
\end{equation}
which is expected to be $\ell=(n+1)$-loop exact. For $\ell=0$ the operator \eqref{LoopOsoft} of course agrees with~\eqref{TreeOsoft}.
In the following we will show that the operators \eqref{LoopOsoft} obey the same type of conservation equations $\Dcal^\pm \Ocal^{\ell-{\rm loop}}_{1-\ell,\pm s}\simeq 0$ as the tower of conformally soft tree operators --- albeit with different contact terms.

\subsection{Conformally soft theorems from trees to loops}

The infrared-regulated integrals \eqref{MassiveKsmomentum} simplify in the massless limit where $p_i^2=m^2 \to 0$, $\beta_{ij}\to 1$ and $\ln\frac{1+\beta_{ij}}{1-\beta_{ij}}\to\ln\frac{4(p_i\cdot p_j)^2}{m^4}$; the sums involving $\ln (m^2)$ terms vanish due to charge and momentum conservation. 
Writing\footnote{Note that because in the soft limit $\omega R\to 1$ and thus $\ln (\omega R) \to 0$ the gravitational drag $e^{\Kcal_{\rm drag}}\to 1$ does not contribute to conformally soft theorems.}
\begin{equation}
    \Kcal_{\rm em}=\ln(\omega L) \kcal_{\rm em},\qquad \Kcal_{\rm gr}=\ln(\omega L) \kcal_{\rm gr},
\end{equation}
we extract the coefficients of the energetically soft logarithms\footnote{Here we will not discuss collinear divergences for $i= j$. In the classical limit this was addressed in~\cite{Choi:2025mzg} where
the relation between the logarithmic soft theorems in massless QED and their asymptotic symmetry interpretation was treated.
}
\begin{equation}\label{kcal}
  \kcal_{\rm em} = \frac{\gamma_{\rm em}}{2}\sum_i \sum_{j\neq i}  Q_i Q_j \,\kcal_{ij},\qquad
   \kcal_{\rm gr}= 
 \gamma_{\rm gr} \sum_i \sum_{j\neq i} (p_i\cdot p_j) \, \kcal_{ij},
\end{equation}
where we introduced $\gamma_{\rm em}=\frac{e^2}{4\pi^2}$ and $\gamma_{\rm gr}=\frac{\kappa^2}{32\pi^2}$ and abbreviated
\begin{equation}
   \kcal_{ij}=i\pi\delta_{\eta_i\eta_j,1}-\ln\left(2|p_i\cdot p_j|\right).
\end{equation}
Note that, while the form of $\Kcal_{\rm em}$ and $\Kcal_{\rm gr}$ is reminiscent of the exponentiated infrared-divergent expressions that can be recast as correlators of Goldstone bosons \cite{Nande:2017dba,Himwich:2020rro,Arkani-Hamed:2020gyp}, the unphysical infrared divergence $\ln \Lambda_{\rm IR}$ there is here replaced by $\ln (\omega L)$ and the exponentiation of the logarithmic soft factors involves the operator $\frac{S_{0,i}}{S_{-1,i}}$ acting on $\Kcal_{\rm em}$ and $\Kcal_{\rm gr}$.
In massless scalar QED the logarithmic soft factors are determined by
\begin{equation}
    \frac{S_{0,i}}{S_{-1,i}}\kcal_{\rm em} = \left(-\gamma_{\rm em}\right) \sum_{j\neq i}\frac{Q_i Q_j}{p_i\cdot p_j}   \left[(q\cdot p_j)-(q\cdot p_i)\frac{\varepsilon\cdot p_j}{\varepsilon\cdot p_i}\right].
\end{equation}
Notice that the (imaginary) classical contribution vanishes, which can be explained from asymptotic symmetries~\cite{Choi:2025mzg}.
In gravity we have
\begin{equation}
    \frac{S_{0,i}}{S_{-1,i}}\kcal_{\rm gr} =\left(-2\gamma_{\rm gr}\right) \sum_{j\neq i}\left(1-\kcal_{ij}\right)\left[(q\cdot p_j)-(q\cdot p_i)\frac{\varepsilon\cdot p_j}{\varepsilon\cdot p_i}\right],
\end{equation}
where $(1-\kcal_{ij})$ accounts for both (imaginary) classical and (real) quantum contributions.
We now recast logarithmic energetically soft theorems as conformally soft celestial amplitudes.

\subsubsection{Scalar QED}

The leading tree-level soft photon factor \eqref{SoftTreeEM} in the parametrization \eqref{pmassless} is given by
 \begin{equation}
S^{\rm em,+}_{-1,i}=\sqrt{2}e\frac{Q_i}{z-z_i}
\end{equation}
while the action of the angular momentum operator on the  infrared-regulated loop integral \eqref{kcal} takes the form
\begin{equation}
   \frac{S_{0,i}^{+}}{S_{-1,i}^{+}}\kcal_{\rm em} = -\gamma_{\rm em} \sum_{j\neq i}(\zb-\zb_j)\frac{Q_i Q_j}{\zb_{ij}} \eta_i e^{-\del_{\Delta_i}}.
\end{equation}

\paragraph{Tree.}

The leading tree-level conformally soft theorem is given by
\begin{equation}\label{EMTreeWI}
\langle \Ocal^{\rm tree}_{1,+1}  \dots \rangle =\sqrt{2}e\sum_i \frac{Q_i}{z-z_i} \langle \cdots \rangle,
\end{equation}
and the $\Delta=1$ conformally soft tree operator is conserved as
\begin{equation}
\delbar \Ocal^{\rm tree}_{1,+1}
\simeq 0,
\end{equation}
which is a type II primary descendant. 
The $\ncal=1$ tower of charges labeled by $m=0$ acts on the $i$-th operator in a local manner as
\begin{equation}\label{QLeadingPhoton}
[Q^{0,n},\Ocal_i]=\sqrt{2}e z_i^n Q_i \Ocal_i.
\end{equation}
The action of the charge on the $i$-th operator is thus given by its symmetry variation, $[Q_{\Sigma_i},\Ocal_i]=\delta \Ocal_i$. 
A similar analysis can be carried out for the tower of subleading tree-level soft photon theorems \cite{Pano:2023slc,Banerjee:2019aoy,Banerjee:2019tam}; they are, however, not universal, unlike the tower of logarithmic soft theorems~\cite{Sahoo-Sen}.

\paragraph{Loop.}

The leading logarithmic conformally soft photon theorem is given by
\begin{equation}\label{1LoopEMcorrelator}
 \langle \Ocal^{\rm 1-loop}_{0,+1} \dots\rangle =
\sqrt{2}e\sum_i \frac{Q_i}{z-z_i} \Big[-\gamma_{\rm em}\sum_{j\neq i} (\zb-\zb_j)\frac{Q_i Q_j}{\zb_{ij}}  \eta_i e^{-\del_{\Delta_i}}\Big] \langle \dots\rangle.
\end{equation}
The $\Delta=0$ conformally soft 1-loop operator is conserved as
\begin{equation}\label{delbar2Ocal01}
\delbar^2\Ocal^{\rm 1-loop}_{0,+1}
\simeq 0,
\end{equation}
which is a type III primary descendant. Note that this is the same type of conservation equation as for the $\Delta=0$ conformally soft tree operator $\Ocal^{\rm tree}_{0,+1}$ but with a different contact term resolving the `$\simeq 0$'. This is consistent with the result in \cite{Choi:2025mzg} that the asymptotic symmetry that underlies the logarithmic soft photon theorem in massless scalar QED is superphaserotation symmetry --- the same symmetry that explains the tree-level subleading soft photon theorem.

At higher loop order, we have 
\begin{equation}\label{AllLoopEMcorrelator}
 \langle \Ocal^{\rm \ell-loop}_{\ell-1,+1} \dots\rangle= 
\sqrt{2}e\sum_i \frac{Q_i}{z-z_i} \frac{1}{\ell!}\Big[-\gamma_{\rm em}\sum_{j\neq i} (\zb-\zb_j)\frac{Q_i Q_j}{\zb_{ij}} \eta_i e^{-\del_{\Delta_i}}\Big]^\ell \langle \dots\rangle.
\end{equation}
with the $\Delta=1-\ell$ conformally soft $\ell>1$-loop operator being conserved as a type I primary descendant \begin{equation}
    \delbar^{1+\ell}\Ocal_{1-\ell,+1}^{\rm \ell-loop}\simeq 0.
\end{equation} 
This is the same type of conservation equation as for the tree-level subleading soft operators $\Ocal^{\rm tree}_{\Delta,+2}$ with $\Delta \leq -1$, but again with different contact terms resolving the `$\simeq 0$'. 
The sum over all universal loop operator insertions exponentiates 
\begin{equation}\label{AllLoopEMcorrelatorEXP}
 \langle \sum_{\ell=0}^\infty\Ocal^{\rm \ell-loop}_{\ell-1,+1} \dots\rangle= 
\sqrt{2}e\sum_i \frac{Q_i}{z-z_i} e^{-\gamma_{\rm em}\sum_{j\neq i} (\zb-\zb_j)\frac{Q_i Q_j}{\zb_{ij}} \eta_i e^{-\del_{\Delta_i}}} \langle \dots\rangle.
\end{equation}

In contrast to tree-level soft theorems where the soft factor involved only sums over individual insertion points of hard matter operators, as in \eqref{EMTreeWI}, logarithmic soft theorems involve an additional sum over pairs of hard operators at 1-loop and multiple such sums at higher loop-order. Notice though that the exponentiated sums always connect {\it pairs} of hard matter operators ($ij$) anchored at the same hard operator ($i$).
In these pairwise sums the soft photon direction, moreover, appears through the distance to the hard matter operator insertions on the celestial sphere. 
The symmetry representation is non-local and the correlators of the conserved conformally soft loop operators are non-meromorphic\footnote{
Correlators with holomorphic behavior may instead be obtained by inserting the celestial photon current, defined as the $\delbar^{1-\Delta}$ descendant of the $J=+1$ conformally soft $\ell$-loop operator \eqref{LoopOsoft}, which yields
\begin{equation}
 \langle \jcal_\ell^{\rm em}(z) \dots\rangle =
\sqrt{2}e\sum_i \frac{Q_i}{z-z_i} \Big[-\gamma_{\rm em}\sum_{j\neq i} \frac{Q_i Q_j}{\zb_{ij}}  \eta_i e^{-\del_{\Delta_i}}\Big]^\ell \langle \dots\rangle .
\end{equation} 
The tree current $\jcal^{\rm em}_{\ell=0}=J(z)$ is the U(1) current generating the superphaserotation symmetry that underlies the leading soft photon theorem.
}.

\subsubsection{Gravity}

The leading tree-level soft graviton factor \eqref{SoftTreeEM} in the parametrization \eqref{pmassless} is given by
\begin{equation}
    S_{-1,i}^{\rm gr,+}=-\frac{\kappa}{2} \frac{\zb-\zb_i}{z-z_i} \eta_i e^{\del_{\Delta_i}}.
\end{equation}
The action of the angular momentum operator on the infrared regulated loop integral \eqref{kcal} is
\begin{equation}\label{S0/Sm1kgr}
    \frac{S_{0,i}^{+}}{S_{-1,i}^{+}}\kcal_{\rm gr} =\gamma_{\rm gr}  \sum_{j\neq i} (\zb-\zb_j)z_{ij}(1-\kcal_{ij}) \eta_j e^{\del_{\Delta_j}}.
\end{equation} 

\paragraph{Tree.}
The leading tree-level conformally soft theorem is given by
\begin{equation}
\langle \Ocal^{\rm tree}_{1,+2}\dots\rangle = -\frac{\kappa}{2}\sum_i \frac{\zb-\zb_i}{z-z_i} \eta_i e^{\del_{\Delta_i}} \langle\dots\rangle.
\end{equation}
The $\Delta=1$ conformally soft tree operator is conserved as
\begin{equation}
\delbar^2 \Ocal^{\rm tree}_{1,+2}
\simeq 0,
\end{equation} 
which is a type II primary descendant. There are $\ncal=2$ towers of charges labeled by $m=0,1$ which act on the $i$-th operator as
\begin{equation}\label{QLeadingGraviton}
\badat{2}
[Q^{0,n},\Ocal_i]&=-\frac{\kappa}{2} z_i^n  \eta_i e^{\del_{\Delta_i}} \Ocal_i,\\
[Q^{1,n},\Ocal_i]&=-\frac{\kappa}{2}  z_i^n \zb_i \eta_i e^{\del_{\Delta_i}} \Ocal_i.
\eadat
\end{equation}
The tree-exact topological charge operator acts on the $i$-th operator in a local way as $[Q_{\Sigma_i},\Ocal_i]=\delta \Ocal_i$. Again, a similar analysis can be carried out for the tower of subleading tree-level soft graviton theorems \cite{Pano:2023slc,Banerjee:2019aoy,Banerjee:2019tam}, which are not universal, though, unlike the tower of logarithmic soft theorems~\cite{Sahoo-Sen}.

\paragraph{Loop.}
The leading logarithmic conformally soft graviton theorem takes the form
\begin{equation}\label{1LoopGRcorrelator}
\langle \Ocal^{\rm 1-loop}_{0,+2} \dots\rangle = -\frac{\kappa}{2}\sum_i \frac{\zb-\zb_i}{z-z_i} \eta_i e^{\del_{\Delta_i}}  \Big[\gamma_{\rm gr}\sum_{j\neq i} (\zb-\zb_j)   z_{ij}(1-\kcal_{ij})  \eta_j e^{\del_{\Delta_j}}\Big]
\langle \dots\rangle.
\end{equation}
The $\Delta=0$ conformally soft 1-loop operator is conserved as
\begin{equation}
\delbar^3 \Ocal^{\rm 1-loop}_{0,+2}
\simeq 0,
\end{equation}
which is a type II primary descendant. This is the same type of conservation equation as for the $\Delta=0$ conformally soft tree operator $\Ocal^{\rm tree}_{0,+2}$ --- albeit with a different contact term resolving the `$\simeq 0$'.\footnote{For gravity coupled to massive matter this is consistent with the result in \cite{Choi:2024ajz} that superrotation symmetry governs both the subleading tree-level as well as the leading logarithmic soft theorem. We expect a similar result for gravity coupled to massless matter, which will be presented elsewhere \cite{Choi-Kadhe-Puhm_wip}.}
At higher loop order, we have 
\begin{equation}\label{AllLoopGRcorrelator}
\badat{2}
\langle \Ocal^{\rm \ell-loop}_{1-\ell,+2} \dots \rangle =-\frac{\kappa}{2}  \sum_i \frac{\zb-\zb_i}{z-z_i}\eta_i e^{\del_{\Delta_i}} \frac{1}{\ell!}\Big[ \gamma_{\rm gr}\sum_{j\neq i} (\zb-\zb_j) z_{ij} (1-\kcal_{ij}) \eta_j e^{\del_{\Delta_j}}\Big]^\ell
\langle \dots\rangle,
\eadat
\end{equation}
with the $\Delta=1-\ell$ conformally soft $\ell$-loop graviton operator being conserved as 
\begin{equation}\label{Primary_descendent condition for l-loop graviton operator}
\delbar^{2+\ell} \Ocal^{\rm \ell-loop}_{1-\ell,+2}\simeq 0.
\end{equation}
For $\ell=2$ this corresponds to a type III primary descendant, and for $\ell>2$ we have a type I primary descendant. This is again the same conservation equation as for the tree-level subleading soft operators $\Ocal^{\rm tree}_{\Delta,+2}$ with $\Delta\leq -2$, albeit with different contact terms resolving the `$\simeq 0$'. 
The sum over all universal loop operator insertions exponentiates
\begin{equation}\label{AllLoopGRcorrelatorEXP}
\badat{2}
\langle \sum_{\ell=0}^\infty\Ocal^{\rm \ell-loop}_{1-\ell,+2} \dots \rangle =-\frac{\kappa}{2}  \sum_i \frac{\zb-\zb_i}{z-z_i}\eta_i e^{\del_{\Delta_i}} e^{\gamma_{\rm gr}\sum_{j\neq i} (\zb-\zb_j) z_{ij} (1-\kcal_{ij}) \eta_j e^{\del_{\Delta_j}}}
\langle \dots\rangle.
\eadat
\end{equation}

Again we see that, in contrast to the tree-level analysis, the symmetry representation is non-local due to the appearance of products of pairwise sums between individual hard operators as well as between the soft graviton operator and hard matter operators. This, moreover, renders the conformally soft correlator non-meromorphic\footnote{To obtain holomorphic correlators we may consider the insertion of the celestial graviton current, defined as the $\delbar^{2-\Delta}$ descendant of the $J=+2$ conformally soft $\ell$-loop operator  \eqref{Loo[Projector}, which takes the form
\begin{equation}
 \langle \jcal_\ell^{\rm gr}(z) \dots\rangle =
-\frac{\kappa}{2}\sum_i \frac{\eta_i P_i}{z-z_i} (\ell+1) \Big[\gamma_{\rm gr}\sum_{j\neq i} z_{ij}(1-\kcal_{ij}) \eta_j P_j\Big]^\ell \langle \dots\rangle.
\end{equation}
The tree current $\jcal^{\rm gr}_{\ell=0}=P(z)$ is the supertranslation current generating BMS symmetry that underlies the leading soft graviton theorem. 
}.

\subsubsection{Einstein-Maxwell}

The logarithmic soft photon and graviton theorems in Einstein-Maxwell theory at $\ell$-loop order are 
\begin{equation}
    \langle \Ocal^{\rm \ell-loop}_{1-\ell,+s}...\rangle = \sum_iS_{-1,i} \frac{1}{\ell!}\Big\{ \sum_{j\neq i} (\zb-\zb_j) \Big[-\gamma_{\rm em}\frac{Q_i Q_j}{\zb_{ij}} \eta_i e^{-\del_{\Delta_i}}+\gamma_{\rm gr}  z_{ij}(1-\kcal_{ij}) \eta_j e^{\del_{\Delta_j}}\Big]\Big\}^\ell
\end{equation}
where $S_{-1,i}$ is the leading soft photon factor for $s=1$ and the leading soft graviton factor for $s=2$.
The $\Delta=1-\ell$ conformally soft $\ell$-loop operator is conserved as 
\begin{equation}
\delbar^{s+\ell} \Ocal^{\rm \ell-loop}_{1-\ell,+s}\simeq 0.
\end{equation}
This is the same type of conservation equation as for the spin-$s$ logarithmic soft theorems in scalar QED and gravity above but with different contact terms.
For $s=1$ the conservation equation corresponds to a type II, III and I primary descendant for, respectively, $\ell=0$, $\ell=1$ and $\ell>1$. Unlike in scalar QED the classical limit no longer vanishes due to gravitational interactions.
For $s=2$ the conservation equation corresponds to a type II primary descendant for $\ell=0$ and $\ell=1$, to a type III primary descendant for $\ell=2$ and to a type I primary descendant for $\ell>2$.
Both conformally soft theorems exponentiate when summed over all loops.

\section{Conformally soft OPEs from trees to loops}
\label{SoftOPEs}

In this section we compute the celestial operator product expansion (OPE) between two conformally soft operators, generalizing the results for the universal tower of tree operators \eqref{TreeOsoft} to the tower of loop operators \eqref{LoopOsoft}.
Recall that the projector for the universal tree tower in energy and boost basis is given by
\begin{equation}\label{TreeProjector}
  \lim_{\omega\to 0}  \del_\omega^{n+1}\,\omega\,(.) \quad \to \quad \lim_{\Delta\to -n} (\Delta+n)\,(.)\equiv \frac{1}{2\pi i} \oint_{\Delta=n}d\Delta\, (.)\;.
\end{equation}
Meanwhile, the projector for the universal logarithmic tower is
\begin{equation}\label{Loo[Projector}
  \lim_{\omega\to 0}  \frac{1}{\ell!}\left(\prod_{k=1}^\ell\frac{(\omega\del_\omega)^k}{k!}\,\del_\omega\,\right)\omega\,(.) \quad \to \quad \lim_{\Delta\to -n} (\Delta+\ell-1)^{\ell+1}\frac{(-1)^\ell}{\ell!}\,(.)\;
  ,
\end{equation}
where the ordering of the operator product is defined from right to left: $\prod_{k=1}^\ell A_k\equiv A_\ell\cdots A_1$.
We will consider the consecutive double soft limit by applying these projectors to the celestial correlator consecutively for the first and second operator that is taken soft. Our focus will be on extracting the singular celestial OPE between universal conformally soft tree and loop operators which is obtained in the (anti)holomorphic collinear limit\footnote{In this section we are treating $z$ and $\zb$ as real and independent.} $z\to z'$ ($\zb\to\zb'$) of the operator insertions at $(z,\zb)$ and $(z',\zb')$ on the celestial sphere\footnote{Note that we previously used a prime to denote SL(2,$\mathbb C$)-covariant polarization vectors which depend on a reference point $(z',\zb')$. In the celestial OPE this reference point drops out and we henceforth drop the primes on polarizations altogether. In abuse of notation we will in this section denote by ($\zb,\zb'$) instead the location of one of the soft operators.}.

\subsection{Tree OPEs}

The tree-level celestial operator product expansion (OPE) between two conformally soft operators can be extracted from the the double soft insertion of the tree operators \eqref{TreeOsoft} \cite{Mitra-Pano-Puhm_toappear}. Here we are interested in extracting the singular OPE from the double soft expansion\footnote{The double soft limit also captures all regular terms~\cite{Mitra-Pano-Puhm_toappear}.} 
\begin{equation}
    \langle \Ocal^{\rm tree}_{-n,\pm s}(z,\zb)  \Ocal^{\rm tree}_{-n',\pm s}(z',\zb')\dots \rangle = \lim_{\Delta\to -n}(\Delta+n) \lim_{\Delta'\to -n'}(\Delta'+n')  \langle \Ocal^{\rm tree}_{\Delta,\pm s}(z,\zb)  \Ocal^{\rm tree}_{\Delta',\pm s}(z',\zb')\dots \rangle.
\end{equation}
For two outgoing positive helicity soft operators this corresponds to the holomorphic collinear limit $z\to z'$. This yields
\begin{equation}\label{SoftTreeOPE}
 \langle \Ocal^{\rm tree}_{-n,+ s}(z,\zb)  \Ocal^{\rm tree}_{-n',+ s}(z',\zb')\dots \rangle \overset{\cdot}{=} \lim_{\Delta'\to -n'}(\Delta'+n') S_{n,'} (z,\zb;z',\zb',\Delta',J') \langle  \Ocal^{\rm tree}_{\Delta',+ s}(z',\zb')\dots \rangle,
\end{equation}
where $\overset{\cdot}{=}$ denotes equality up to regular terms as $z\to z'$. In this limit only the soft factor 
\begin{equation}
S_{n,'}=S_{-1,'} \frac{1}{(n+1)!}\left(\frac{S_{-1,'}}{S_{0,'}}\right)^{n+1},
\end{equation}
associated with the operator $\Ocal^{\rm tree}_{\Delta',+s}(z',\zb')$, contributes.
Note that the singular OPE \eqref {SoftTreeOPE} corresponds to the residue of the contour integral
\begin{equation}\label{TreeSoftOPEResidue}
 \langle \Ocal^{\rm tree}_{-n,\pm s}(z,\zb)  \Ocal^{\rm tree}_{-n',\pm s}(z',\zb')\dots \rangle \overset{\cdot}{=}\frac{1}{2\pi i}\oint_{\Delta'=-n'} \,d\Delta'\;S_{n,'}(z,\zb;z',\zb',\Delta',J')\; \langle  \Ocal^{\rm tree}_{\Delta',\pm s}(z',\zb')\dots \rangle
\end{equation}
around the simple conformally soft pole at $\Delta'=-n'\in 1-\mathbb{Z}_{\geq0}$.

\subsubsection{Scalar QED}

Since photons are uncharged we have
\begin{equation}
\Ocal^{\rm tree}_{-n,+ 1}(z,\zb)  \Ocal^{\rm tree}_{-n',+ 1}(z',\zb')  \overset{\cdot}{=} 0,
\end{equation}
and so all conformally soft tree OPEs vanish.

\subsubsection{Gravity}
In gravity, with the tree-level soft factor 
\begin{equation}
S^{+\rm gr}_{n,'}= 
-\frac{\kappa}{2}\frac{\zb-\zb'}{z-z'} e^{\del_{\Delta'}}\frac{1}{(n+1)!}\left[ e^{-\del_{\Delta'}}\left(-\Delta'+J'+(\zb-\zb')\del_{\zb'}\right)\right]^{n+1},
\end{equation}
acting on the correlator $\langle \dots \rangle$,
we can extract from the double soft limit \eqref{TreeSoftOPEResidue}
the OPE between two conformally soft graviton operators \eqref{LoopOsoft} which is given by~\cite{Himwich:2023njb,Mitra-Pano-Puhm_toappear}
\begin{equation}
\Ocal^{\rm tree}_{-n,+2}(z,\zb)\Ocal^{\rm tree}_{-n',+2}(z',\zb')\overset{\cdot}{=}	-\frac{\kappa}{2}\frac{1}{z-z'}\sum_{m=0}^{n+1} \frac{1}{m!} \binom{n+n'+2-m}{n'+1} (\zb-\zb')^{m+1} \del_{\zb'}^m \Ocal_{-n-n',2}(z',\zb').
\end{equation}
Expanding the conformally soft operators as in \eqref{Osoftexpansion}, it was shown that the commutator of the associated charges \eqref{Qmn}, after a light-transform, forms a closed algebra known as $w_{1+\infty}$.

\paragraph{BMS OPEs.}
The BMS algebra, consisting of supertranslations and superrotations, forms a closed subalgebra of $w_{1+\infty}$.  This can be inferred from the OPEs of the $\Delta=1$ leading soft graviton operator generating supertranslations and the $\Delta=0$ subleading soft graviton operator generating superrotations.
For two leading soft graviton operators
\begin{equation}\label{STwSROPE}
  \Ocal^{\rm tree}_{1,+2}(z,\zb)\Ocal^{\rm tree }_{1,+2}(z',\zb') \overset{\cdot}{=} -\frac{\kappa}{2} \frac{\zb-\zb'}{z-z'} \Ocal^{\rm tree}_{2,+2}(z',\zb'),
\end{equation}
 where we allowed for the possibility of a double pole $\sim \frac{1}{\omega^2}$ in the double soft limit associated to the operator $\Ocal^{\rm tree}_{2,+2}(z',\zb')$; this corresponds to the $1$ in $w_{1+\infty}$ which is a central term. In the absence of such a double pole the OPE between two leading soft gravitons operators vanishes in line with supertranslations forming an abelian algebra.
The OPE between the leading and subleading soft graviton operator is given by
\begin{equation}
\Ocal^{\rm tree}_{0,+2}(z,\zb)\Ocal^{\rm tree}_{1,+2}(z',\zb') \overset{\cdot}{=} -\frac{\kappa}{2} \frac{\zb-\zb'}{z-z'} \Ocal^{\rm tree}_{1,+2}(z',\zb').
\end{equation}
This is consistent with the commutator between superrotations and supertranslations yielding another supertranslation. 
For two subleading soft graviton operators the OPE is
 \begin{equation}
        \Ocal^{\rm tree}_{0,+2}(z,\zb)\Ocal^{\rm tree}_{0,+2}(z',\zb') \overset{\cdot}{=} -\frac{\kappa}{2} \frac{\zb-\zb'}{z-z'} \left[-2+(\zb'-\zb)\del_{\zb'}\right] \Ocal^{\rm tree}_{0,+2}(z',\zb'),
    \end{equation}
which agrees with the commutator of two superrotations giving another superrotation.

\subsection{Loop OPEs}

We extract the loop-level celestial OPE  between two conformally soft operators from the double soft insertion of the loop operators \eqref{LoopOsoft} given by
\begin{equation}
\badat{2}
    &\langle \Ocal^{\rm \ell-loop}_{-1-\ell,\pm s}(z,\zb)  \Ocal^{\rm \ell'-loop}_{1-\ell',\pm s}(z',\zb')\dots \rangle \\
    &= \lim_{\Delta\to 1-\ell}\frac{(-1)^\ell}{\ell!}(\Delta+\ell-1)^{\ell+1} \lim_{\Delta'\to 1-\ell'}\frac{(-1)^{\ell'}}{{\ell'}!}(\Delta'+\ell'-1)^{\ell'+1} \langle \Ocal^{\rm \ell-loop}_{\Delta,\pm s}(z,\zb)  \Ocal^{\rm \ell'-loop}_{\Delta',\pm s}(z',\zb')\dots \rangle.
\eadat
\end{equation}
Again we are interested only in the singular OPE which for outgoing positive helicity soft operators is obtained from the holomorphic collinear limit $z\to z'$ of the double soft insertion  
\begin{equation}\label{SoftLoopOPE}
\badat{2}
& \langle \Ocal^{\rm \ell-loop}_{1-\ell,+ s}(z,\zb)  \Ocal^{\rm \ell'-loop}_{1-\ell',+ s}(z',\zb')\dots \rangle \\
&\quad\quad \overset{\cdot}{=} \lim_{\Delta'\to 1-\ell'}\frac{(-1)^\ell}{\ell!}(\Delta'+\ell'-1)^{\ell'+1} S^{\ln}_{n,'} (z,\zb;z',\zb',\Delta',J') \langle  \Ocal^{\rm \ell'-loop}_{\Delta',+ s}(z',\zb')\dots \rangle,
\eadat
\end{equation}
where $\overset{\cdot}{=}$ denotes equality up to terms that are regular as $z\to z'$. The only singular contribution arises from the logarithmic soft factor 
\begin{equation}
S^{\ln}_{1-\ell,'}=S_{-1,'} \frac{1}{\ell!}\left(\frac{S_{-1,'}}{S_{0,'}}\kcal\right)^{\ell},
\end{equation}
associated with the operator $\Ocal^{\rm tree}_{\Delta',+s}(z',\zb')$.

\subsubsection{Scalar QED}
The OPE of the conformally soft photon loop operators again vanishes,
\begin{equation}
\label{OellOellprimesQED}
\Ocal^{\rm \ell-loop}_{1-\ell,+ 1}(z,\zb)  \Ocal^{\rm \ell'-loop}_{1-\ell',+ 1}(z',\zb')  \overset{\cdot}{=} 0,
\end{equation}
since photons are uncharged.

\subsubsection{Gravity}

Recall that in gravity the logarithmic $\ell$-loop soft factor for the universal tower with $\ell=n+1$ is
\begin{equation}
    S^{\ln+\rm gr}_{1-\ell,'}=-\frac{\kappa}{2} \frac{\zb-\zb'}{z-z'} e^{\del_{\Delta'}}\frac{1}{
\ell!}\Big[\gamma_{\rm gr} \sum_{j}(\zb-\zb_j)(z'-z_j)(1-\kcal_{'j})\eta_j e^{\del_{\Delta_j}}\Big]^{\ell}.
\end{equation}
Plugging this into \eqref{SoftLoopOPE} we note that 
\begin{equation}
    \lim_{\Delta'\to 1-\ell'}(\Delta'+\ell'-1)^{\ell'+1} e^{\del_{\Delta'}} \langle  \Ocal_{\Delta',+ 2}(z',\zb')\dots \rangle=0,
\end{equation}
since the insertion of $e^{\del_{\Delta'}} \Ocal_{\Delta',+ 2}= \Ocal_{\Delta'+1,+ 2}$ does not give rise to poles at $\Delta'= 1-\ell'$ of order $\ell'+1$ and so is killed by the projector.
Hence the celestial OPE between two conformally soft graviton loop operators \eqref{LoopOsoft} vanishes,
\begin{equation}
\label{OellOellprimeGR0}
    \Ocal^{\rm \ell-loop}_{1-\ell,+ s}(z,\zb)  \Ocal^{\rm \ell'-loop}_{1-\ell',+ s}(z',\zb') \overset{\cdot}{=} 0.
\end{equation}

Our focus here was on computing the celestial OPE for the loop operators \eqref{LoopOsoft} which give rise to universal logarithmic soft factors. We may also be interested in the OPE of these operators with the tower of tree-level operators or with non-universal conformally soft loop operators, both of which are corrected at higher loops and by effective field theory operators. We will conclude this section by discussing the loop corrections to the tree-level BMS OPEs. 

\paragraph{Loop-corrected BMS OPEs.}
While the $\Delta=1$ leading soft graviton operator is tree-exact, the $\Delta=0$ subleading soft graviton operator is corrected at one-loop. We now investigate how this affects the tree-level BMS OPEs. 

From \eqref{OellOellprimeGR0} it follows that the celestial OPE of two subleading soft graviton operators is
\begin{equation}
\Ocal^{\rm 1-loop }_{0,+2}(z,\zb)\Ocal^{\rm 1-loop }_{0,+2}(z',\zb') \overset{\cdot}{=} 0.
\end{equation}
To compute the celestial OPE between the leading soft tree graviton operator and the subleading soft loop graviton operator from the double soft limit there are two possible orderings. 
Applying first the $\Delta\to 1$ tree projector followed by the $\Delta'\to 0$ loop projector yields 
\begin{equation}
\badat{2}
\langle\Ocal^{\rm 1-loop }_{0,+2}(z,\zb)\Ocal^{\rm tree}_{1,+2}(z',\zb') \dots \rangle&\overset{\cdot}{=} \lim_{\Delta'\to0} \left(-{\Delta'}^2\right) \left[-\frac{\kappa}{2} \frac{\zb-\zb'}{z-z'} e^{\del_{\Delta'}}\right]\langle \Ocal_{\Delta',+2}\dots\rangle \overset{\cdot}{=}0,
\eadat
\end{equation}
which vanishes because $e^{\del_{\Delta'}}\Ocal_{\Delta',+2}=\Ocal_{\Delta'+1,+2}$ does not have a double pole at $\Delta'=0$. 
 Thus we find that the celestial OPE
\begin{equation}\label{O01loopO1tree}
    \Ocal^{\rm 1-loop}_{0,+2}(z,\zb)\Ocal^{\rm tree}_{1,+2}(z',\zb') \overset{\cdot}{=}0.
\end{equation}
Taking the double soft limit in the opposite order, applying first the $\Delta\to 0$ loop projector followed by the $\Delta'\to 1$ tree projector, yields
\begin{equation}
\badat{2}
&\Ocal^{\rm tree}_{1,+2}(z,\zb)\Ocal^{\rm 1-loop}_{0,+2}(z',\zb') \\
&\qquad \overset{\cdot}{=}\frac{1}{2\pi i} \oint_{\Delta'= 1} d\Delta'\,\left[-\frac{\kappa}{2} \frac{\zb-\zb'}{z-z'} e^{\del_{\Delta'}}\right] \left[\gamma_{\rm gr} \sum_j(\zb-\zb_j)(z'-z_j)(1-\kcal_{'j})\eta_j e^{\del_{\Delta_j}}\right] \langle \Ocal_{\Delta',+2}\dots\rangle\\
&\qquad\overset{\cdot}{=} -\frac{\kappa}{2} \frac{\zb-\zb'}{z-z'} \gamma_{\rm gr} \sum_j(\zb-\zb_j)(z'-z_j)(1-\kcal_{'j})\eta_j e^{\del_{\Delta_j}}\langle \Ocal^{\rm tree}_{2,+2}\dots\rangle.
\eadat
\end{equation}
Note that here we use as tree projector the contour integral in \eqref{TreeProjector} rather than the multiplicative projector which is not well-defined due to the higher-order poles generated by loops. After shifting the contour variable to $\Delta''=\Delta'+1$ we pick up a simple pole at $\Delta''=2$. This corresponds to the operator $\Ocal^{\rm tree}_{2,+2}$ which, as discussed at for tree-level OPEs vanishes in the absence of double poles $\sim \frac{1}{\omega^2}$ in the double soft limit. The OPE we extract from this order of soft limits then agrees with \eqref{O01loopO1tree}.

The remaining OPE is between the subleading soft tree and loop graviton operators. Extracting these OPEs from the double soft limit there are again two possible orderings.
Applying first the $\Delta\to 0$ tree projector before applying the $\Delta\to 0$ loop projector yields 
\begin{equation}
\badat{2}
& \Ocal^{\rm tree}_{0,+2}(z,\zb)\Ocal^{\rm 1-loop }_{0,+2}(z',\zb') \\
        &\qquad \overset{\cdot}{=}\lim_{\Delta'\to 0} \left(-{\Delta'}^{2}\right)\left[-\frac{\kappa}{2} \frac{\zb-\zb'}{z-z'} e^{\del_{\Delta'}}\right]\left[e^{-\del_{\Delta'}}\left(\Delta' -J'+(\zb'-\zb)\del_{\zb'}\right) \right]\langle \Ocal_{\Delta'}(z',\zb')\dots \rangle\\
        &\qquad \overset{\cdot}=-\frac{\kappa}{2} \frac{\zb-\zb'}{z-z'}\left[2-(\zb'-\zb)\del_{\zb'}\right]\lim_{\Delta'\to 0} {\Delta'}^{2}\langle \Ocal_{\Delta'}(z',\zb')\dots \rangle.
\eadat
\end{equation}
The operator $\Ocal_{\Delta'}(z',\zb)$ gives rise to a double pole at $\Delta'=0$ at $\ell'=1$ loop. Thus the celestial OPE is  
\begin{equation}\label{O0treeO01loop}
  \Ocal^{\rm tree}_{0,+2}(z,\zb)\Ocal^{\rm 1-loop }_{0,+2}(z',\zb') \overset{\cdot}{=} \frac{\kappa}{2} \frac{\zb-\zb'}{z-z'} \left[2-(\zb'-\zb)\del_{\zb'}\right] \Ocal^{\rm 1-loop}_{0,+2}(z',\zb').
\end{equation}
This reproduces the OPE \eqref{STwSROPE} between the leading and subleading tree-level soft graviton operators except with the subleading tree operator replaced by the corresponding loop operator.

In the opposite soft order, applying first the $\Delta\to 0$ loop projector before taking the $\Delta'\to 0$ tree projector, yields
\begin{equation}
\badat{2}
&\Ocal^{\rm 1-loop }_{0,+2}  (z,\zb)\Ocal^{\rm tree}_{0,+2}(z',\zb')  \\
&\qquad \overset{\cdot}{=} \frac{1}{2\pi i} \oint_{\Delta'= 0}d\Delta'\,  \left[-\frac{\kappa}{2} \frac{\zb-\zb'}{z-z'} e^{\del_{\Delta'}}\right] \left[\gamma_{\rm gr} \sum_j(\zb-\zb_j)(z'-z_j)(1-\kcal_{'j})\eta_j e^{\del_{\Delta_j}}\right] \langle \Ocal_{\Delta',+2}\dots\rangle \\
&\qquad \overset{\cdot}{=}-\frac{\kappa}{2} \frac{\zb-\zb'}{z-z'}  \left[\gamma_{\rm gr} \sum_j(\zb-\zb_j)(z'-z_j)(1-\kcal_{'j})\eta_j e^{\del_{\Delta_j}}\right]   \langle \Ocal^{\rm tree}_{1,+2}\dots\rangle
.
\eadat
\end{equation}
Here we again shifted the contour integration variable $\Delta''=\Delta'+1$ and picking up the residue of the simple pole at $\Delta''=1$ corresponds to the insertion of the leading soft graviton operator $\Ocal^{\rm tree}_{1,+2}$. It appears that the consecutive double soft limit of the leading logarithmic and subleading tree soft operator does not commute --- unlike the situation at tree-level \cite{Mitra-Pano-Puhm_toappear}.

\section{Discussion}
\label{discussion}

The S-matrix in theories with massless particles is infrared-divergent. An infrared-finite observable can be extracted from the ratio of the amplitudes
with $N$ hard particles and an extra soft particle and the amplitude with $N$ hard particles. In the energetically soft limit \cite{Sahoo-Sen} this ratio is proportional to $\omega^n (\ln \omega)^\ell$ where the theory-independent and $\ell$-loop exact terms have $n=\ell-1$. Two different scales appear that make the logarithm dimensionless: the typical system size $L$ and an infrared scale associated to the distance to the detector $R$. We have argued in section \ref{SoftFactSmatrix} that in a Faddeev-Kulish dressed Hilbert~space, poles in graviton loop integrands below the energy scale $1/R$ are canceled and poles from real graviton radiation below the energy scale $1/R$ are screened. Hence, the energetically soft expansion is in the small parameter $\omega L \ll 1$ around $\omega =1/R$, and thus the actual logarithmically soft terms are $\sim \omega^n(\ln \omega L)^\ell$, while the gravitational drag $\sim \omega^n(\ln \omega R)^\ell$ is on a fundamentally different footing.
We leave a more detailed derivation of this for future work.

For the conjectured tower of logarithmic soft theorems with $n=\ell-1$, we argued in section \ref{SoftFactSmatrix} that not all terms in the soft factors~\eqref{SoftTowerBEM} and~\eqref{SoftTowerCGR} contribute to the soft photon and graviton Ward identities, and that those that contribute exponentiate when summing over the universal soft terms at all loops. This soft exponentiation argument was based on the general structure of the classical logarithmic soft photon and graviton theorems.  It would be very interesting to verify if this structure persists for the quantum logarithmic soft factors beyond one-loop.
In massless scalar QED, where the classical logarithmic soft factor vanishes, we assume that the structure~\eqref{SoftTowerBEM} holds beyond one loop when writing the celestial Ward identities in section \ref{CelestialWID}.
For gravity our results of sections \ref{CelestialWID} and \ref{SoftOPEs} account for both classical and quantum logarithmic soft factors in the form of~\eqref{SoftTowerCGR}. 
While the exponentiation of the symmetry-governed part of the logarithmic soft factors applies to both massless and massive particles, in the discussion of the celestial Ward identities we restricted ourselves to massless matter. 
The extension of this to theories with massive particles is work in progress~\cite{Choi-Kadhe-Puhm_massive}.

The main take-away from our discussion in section \ref{CelestialWID} of celestial Ward identities for logarithmic soft theorems in theories with massless matter is as follows. We showed that conformally soft loop operators satisfy the same type of conservation equations as their tree-level counterparts of the same conformal dimension, albeit with different contact terms. This justifies the exponentiation of the symmetry-governed logarithmic tower in section \ref{SoftFactSmatrix}. Unlike at tree level, the loop-corrected celestial Ward identities involve sums over all hard operators and the resulting symmetry action is non-local. The residue that is picked up by a contour integral around a single hard operator insertion at $(z_i,\zb_i)$ involves ($\ell$ products of) pairwise sums $\sum_{j\neq i}$ of hard operator data anchored to the hard operator at $(z_i,\zb_i)$. Moreover, the soft operator location $(z,\zb)$ also enters in the ($\ell$ products of) pairwise sums through the distance to the hard operator at $(z_j,\zb_j)$. Defining instead celestial loop currents removes that dependence on the soft operators location in the sums and gives rise to correlators with holomorphic behavior.

In section \ref{SoftOPEs} we computed the all-order (singular) celestial OPEs between the universal conformally soft loop operators in scalar QED and gravity and show that they vanish. This is one of the main results of this paper. We then study loop corrections to the tree-level OPEs involving operators with conformal dimensions $\Delta=1$ and $\Delta=0$ that give rise to the BMS algebra. Unlike at tree level, where both the double soft graviton limit and the order of taking holomorphic collinear and soft limits commute \cite{Mitra-Pano-Puhm_toappear}, at loop level the situation is more nuanced and depends on whether tree or loop operators are taken soft first.
When the tree operators are taken soft first, we recover the conformally soft OPEs that give rise to the BMS algebra but with the conformally soft tree operator with  $\Delta=0$ (which receives loop corrections) replaced by the conformally soft loop operator with $\Delta=0$ (which is 1-loop exact). 

Our findings hint that the study of the corrections to the tree-level $w_{1+\infty}$ algebra in Einstein gravity to all loop orders requires knowledge beyond the universal logarithmic soft tower $\ell=n+1$, potentially of all non-universal soft factors between $0\leq \ell\leq n+1$.

\section*{Acknowledgments}

We would like to thank Fernando Alday, Lucía Gomez-Córdova, Diego Hofman, Alok Laddha, Prahar Mitra, and Emilio Trevisani for useful discussions. SC and AP are supported by the European Research Council (ERC) under the European Union’s Horizon 2020 research and innovation programme (grant agreement No 852386). This work was supported by the Simons Collaboration on Celestial Holography. \\

\printbibliography

\end{document}